\documentclass[11pt]{article}
\usepackage[utf8x]{inputenc}
\usepackage{amsfonts,amsthm}
\usepackage{amsmath}
\usepackage{complexity}
\usepackage{fullpage}
\usepackage{color}
\usepackage{graphicx}
\usepackage{balance}
 \usepackage{amsthm}
\usepackage{graphicx}

\newtheorem{lemma}{Lemma}
\newtheorem{theorem}{Theorem}

\newtheorem{corollary}{Corollary}

\newtheorem{fact}{Fact}
 \newtheorem*{assumption:unique-prices}{Non-Unique Prices Assumption}
 \newtheorem*{assumption:uniform-prices}{Big-Small Items Assumption}
 \newtheorem*{restriction:unique-prices}{Unique Prices Restriction}
 \newtheorem*{assumption:val-small}{Valuable Small Items Assumption}
 \newtheorem*{assumption:not-full-items}{Not Fully Assigned Items Assumption}
 \newtheorem*{restriction:val-small}{Cheap Small Items Restriction}
 \newtheorem*{restriction:full-items}{Fully Assigned Items Restriction}

\newcounter{restatecounter:lem}
\newcounter{restatecounter:restriction}
\newtheorem{re:lemma}[restatecounter:lem]{Lemma}
\newtheorem{re:theorem}{Theorem}
\newtheorem{re:definition}{Definition}
\newtheorem{re:assumption}{Assumption}
\newtheorem{re:corollary}{Corollary}
\newtheorem{re:claim}{Claim}
\newtheorem{re:restriction}[restatecounter:restriction]{Restriction}

\newcommand{\hide}[1]{}

\newcommand{\players}{\ensuremath{\mathcal{A}}}
\newcommand{\items}{\ensuremath{\mathcal{Q}}}
\newcommand{\conf}{\ensuremath{\mathcal{C}}}
\newcommand{\bigitems}{\ensuremath{\mathcal{B}}}
\newcommand{\smallitems}{\ensuremath{\mathcal{S}}}
\newcommand{\Val}{\ensuremath{\mathsf{Val}}}

\newcommand{\Deltah}{\hat{\Delta}}

\title{An Improved Approximation Guarantee for the Maximum Budgeted Allocation Problem}
\author{Christos Kalaitzis\footnote{School of Computer and Communication Sciences, EPFL. Email:  christos.kalaitzis@epfl.ch.  Supported by
ERC Starting Grant 335288-OptApprox.}}
\date{}

\begin{document}
\maketitle

\begin{abstract}
We study the Maximum Budgeted Allocation problem, which is the problem of assigning indivisible items
 to players with budget constraints. In its most general form, an instance of the MBA problem might include many different prices for the 
 same item among different players, and different budget constraints for every player. So far, the best approximation algorithms we know
for the MBA problem achieve a $3/4$-approximation ratio, and employ a natural LP relaxation, called the Assignment-LP. In this paper, we
give an algorithm for MBA, and prove that it achieves a $3/4+c$-approximation ratio, for some constant $c>0$. This algorithm works by
rounding solutions to an LP called the Configuration-LP, therefore also showing that the Configuration-LP is strictly stronger than the
Assignment-LP (for which we know that the integrality gap is $3/4$) for the MBA problem.
\end{abstract}
\section{Introduction}
Consider the following scenario: you are in charge of the advertisement allocation system of a
website. The website's sponsors are willing to pay certain amounts of money in order to have their
products advertised, and they have announced a predefined budget, i.e., a certain cost threshold
they do not want to exceed. On the other hand, the website has a specified set of advertisement
placements that you are allowed to use. Of course, the attractiveness of these placement options is
not the same, and this is reflected in the fact that the sponsors have announced a maximum amount
of money they are willing to spend for each of these options. Note that
two different sponsors might be willing to pay different amounts of money for the same advertisement
placement. As is natural, you would like to maximize the total revenue you receive from placing the
sponsors' advertisements. This problem 
will be the focus of our work. 

The above scenario can be modeled by the Maximum Budgeted Allocation (MBA) problem, in which there is a
set of agents, each with a specified budget, and a set of indivisible items. 
Every agent has a specific valuation for each item, which represents the maximum amount of money that agent
is willing to spend on each item; however, an agent is not allowed to spend more money
than his budget. The problem is to find
an assignment of items to players, along with corresponding prices, which maximizes the total amount of money the agents pay, without violating 
any agent's budget. Since, given an allocation of items to players, computing optimal prices that respect the agents' budgets is easy, it becomes clear
that the most difficult aspect of the problem is finding such an allocation.
% \sout{Every
% agent has a specified price that he is willing to pay for every item, but he is not allowed to spend more money
% than his budget. The problem is to find
% an assignment of items to players which maximizes the total amount of money the agents pay.} 
The MBA
problem is a generic allocation problem, which has received significant attention from the research
community. The reason is that it arises naturally in the context of multiple practical scenarios, such as when we want
to maximize the revenue in an auction with budget-constrained agents.
However, it is an NP-hard
problem, and therefore hard to solve optimally. Therefore, one natural approach is to take up approximation algorithms
in the course of tackling it.

So far, the best algorithms we know achieve a $3/4$-approximation ratio and are due to Srinivasan \cite{DBLP:conf/approx/Srinivasan08}
and Chakrabarty and Goel \cite{DBLP:journals/siamcomp/ChakrabartyG10}. These algorithms work by rounding solutions to the
Assignment-LP relaxation, a relaxation for which we know the integrality gap is no better than $3/4$. Since our goal is to improve upon the
$3/4$ ratio, we will have to use a stronger relaxation. Configuration LP-s have often been used in the past to provide good approximation 
algorithms (e.g., for the Bin Packing problem \cite{DBLP:conf/focs/KarmarkarK82,DBLP:conf/focs/Rothvoss13}); towards such an end, Chakrabarty and Goel \cite{DBLP:journals/siamcomp/ChakrabartyG10} proposed the use
of the Configuration-LP for the MBA problem, and conjectured that its integrality gap should be better than $3/4$.

At this point, it was already observed \cite{DBLP:journals/siamcomp/ChakrabartyG10} that the 
restriction of items having uniform prices (i.e., an item can be assigned to a subset of the players, and every player is willing to pay the
same price for that item) is one of the most natural restrictions of MBA that do not make the problem too easy. This is in line, for
example, with what was already observed for
the Generalized Assignment Problem (GAP), i.e., the problem of finding an allocation of items to bins of size 1 which maximizes the total
value, where every item has a separate size and value for each bin. Indeed, in their work on designing an algorithm with a
better than $1-1/e$-approximation ratio for GAP, Feige and Vondr\'{a}k \cite{DBLP:conf/focs/FeigeV06} show how to deal with uniform
instances, and then focus on the rest; interestingly, their techniques rely on the use of the
Configuration-LP. Furthermore, these facts highlight some structural differences between different allocation problems:
while problems such as GAP and MBA might not get substantially harder when the item values are
non-uniform, this is not necessarily the case for other allocation problems, as is at least demonstrated by how well the
Configuration-LP performs on uniform and non-uniform instances of problems such as scheduling to minimize the maximum makespan
\cite{DBLP:journals/siamcomp/Svensson12,DBLP:journals/scheduling/VerschaeW14} and maximize the minimum makespan
\cite{DBLP:journals/talg/AsadpourFS12,DBLP:conf/stoc/BansalS06}.

Dealing with the instances with uniform prices will be the main technical challenge we face in
this paper. Therefore, the following question naturally arises: why are these instances the hardest ones? First of all, 
this fact is already indicated by the observation that the worst integrality gap instances for both the Assignment-LP and the
Configuration-LP have uniform prices. However, a deeper reason is the following: with our current techniques, we have a good understanding
of how to manipulate instances that display non-uniform prices in order to exploit this structure and design improved algorithms. However,
when prices are uniform, this is not possible, and making good decisions concerning the assignment of items to players becomes
increasingly important.

\paragraph{Previous work}As we have already stated, the MBA problem is known to be
NP-hard \cite{DBLP:journals/jacm/GandhiKPS06,DBLP:journals/geb/LehmannLN06}, and even NP-hard to
approximate \cite{DBLP:journals/siamcomp/ChakrabartyG10}.
The first approximation algorithm is due to Garg, Kumar and Pandit \cite{garg2001approximation}, and
achieves an approximation guarantee of $\frac{2}{1+\sqrt{5}}$. After this work, there was a series
of improvements (Andelman and Mansour \cite{DBLP:conf/swat/AndelmanM04} provide an algorithm with a guarantee of
$1-1/e$, which was then pushed to $2/3$ by Azar et al. \cite{DBLP:conf/icalp/AzarBKMN08}) which eventually
led to achieving an approximation guarantee of $3/4$, due to
Srinivasan \cite{DBLP:conf/approx/Srinivasan08} and independently due to Chakrabarty and
Goel \cite{DBLP:journals/siamcomp/ChakrabartyG10}. This is the best currently known guarantee;
furthermore, these two works also imply that the integrality gap of the Assignment-LP is
exactly $3/4$. Hence, it becomes clear that in order to achieve something better, we will have to
use a stronger LP relaxation. 

In this direction, the Configuration-LP was employed by Kalaitzis et al.
\cite{DBLP:conf/ipco/KalaitzisMNPS14} to provide a $3/4+c$-approximation algorithm (for some small constant
$c>0$) for a restricted version of the MBA problem (which they call Restricted MBA), in which all the
players have uniform budgets, and the items have uniform prices. The use of the
Configuration-LP for the MBA problem was first proposed by Chakrabarty and Goel
\cite{DBLP:journals/siamcomp/ChakrabartyG10}. The worst known upper bound on the integrality gap of the Configuration-LP is  
$2(\sqrt{2}-1)$\cite{DBLP:conf/ipco/KalaitzisMNPS14}.

\paragraph{Our contributions}Remember that, for the Restricted MBA problem,
we already know an approximation algorithm with a guarantee strictly better than $3/4$ \cite{DBLP:conf/ipco/KalaitzisMNPS14}.
Having this result at hand, a natural way to attack the MBA problem is to try to prove that designing an algorithm with a similar guarantee is possible for any
family of instances that does not fall into this restriction. Carefully choosing which family of instances we will handle at each step,
and providing good approximation algorithms for each one of them,
we will be able to gradually take care of all families of instances that do not fall into the Restricted MBA family, at which point our
work will have been completed.

Formally, our main result is the following:
\begin{theorem}\label{thm:main}
 There is a polynomial time $3/4+c$-approximation algorithm for the MBA problem, for some constant $c>0$;
furthermore, the algorithm establishes the same lower bound on the integrality gap of the
Configuration-LP.
\end{theorem}

In order to achieve this result, we design a new technique to exploit the structure of the family of instances
we examine at times. This technique is an analysis of a randomized version of the well-known rounding algorithm for the
Assignment-LP, which was introduced by Shmoys and Tardos  \cite{DBLP:journals/mp/ShmoysT93} for the problem of scheduling jobs with costs
subject to makespan constraints. The key concept behind
this analysis is what we call {\em worst-case arrangements of configurations}: what we do is examine the structure of the
distribution of the integral solutions this algorithm outputs. Focusing on one player, we look at the distribution of sets of items
this player will receive according to our rounding algorithm, and locate what is the worst-possible distribution the rounding algorithm
could return. Using these ideas, we are able to derive a guarantee on the expected returned value, which not only depends on the
contribution of each player to the LP value, but also quantifies how well the rounding algorithm performs {\em as a function of the
structure} of the fractional solution (e.g., we might be able to obtain different guarantees on the expected value of the sets assigned to two
different players by the rounding algorithm, even though their contribution to the LP value is the same).

Using these ideas, we are able to provide improved approximation guarantees for large families of fractional solutions, until we are
left only with instances where the items have uniform prices, but the agents have non-uniform budgets. In
order to provide strong approximation guarantees for fractional solutions of this family that {\em do not} fall into the Restricted MBA
category (as we mentioned, these solutions we know how to round), we combine our analysis of the Shmoys-Tardos rounding
algorithm with one more idea. The aforementioned analysis indicates the following path: since our guarantee for
the Shmoys-Tardos rounding algorithm depends on the structure of the LP solution, we will try to shift the fractional
assignment of items from one player to the other, and then apply the rounding algorithm, hoping to improve the structure 
of the fractional solution, and thus increase the expected returned value.
The main concept that arises in this context, is that of {\em gain-amplifying players}, i.e., 
players on which the
structure of the LP solution will be favorable to us. Selecting these players will be based on a random partitioning of 
the players. This partitioning defines along which direction items will be fractionally shifted. After this shifting is completed, we
show that the LP solution will have a favorable structure for a subset of the players. This allows us to prove that we can get an improved
approximation guarantee.

\section{Preliminaries}
Let us now define the MBA problem formally, introduce some basic notation, and define
the two main LP relaxations that we will use.
Let $\players$ be the set of players and $\items$ the set of items. Let $B_i$ be the budget of player $i$, and
$p_{ij}$ the price of item $j$ for player $i$. The MBA problem comes down
to finding an indivisible allocation of items to players, i.e., disjoint sets $T_i$ for every $i\in\players$, which
maximize

$$\sum_{i\in\players}\min\left \{\sum_{j\in T_i}p_{ij},B_i\right\}$$

For reasons that will become clear later on, a strategy we  employ
is classifying the items into big and small, depending on their
value relative to some player's budget. Handling these items in a different way is crucial for
our whole approach. We define $\bigitems_i=\{j:p_{ij}\geq (1-\beta)B_i\}$, for some $\beta\leq
1/3$ (which will be set later on) to be the set of big items for player $i$. Also, let $\smallitems_i=\items\setminus \bigitems_i$ be
the set of small items. Sometimes, if $i$ is clear from context, we might drop the subscript.

The first LP relaxation we will use is the Assignment-LP. This formulation
defines a variable $x_{ij}$ for every player $i$ and item $j$; in an integer solution, $x_{ij}$
would be 1 if $j$ is assigned to $i$, and 0 otherwise.

\[
\begin{array}{rrll}
\max & \sum_{i\in \players} P_i&\\
\text{s.t.} & P_i&\leq B_i &\quad \forall i\in \players \\
& P_i&\leq \sum_{j\in \items} x_{ij}p_{ij} &\quad \forall i\in \players \\
& \sum_{i\in \players} x_{ij}&\leq1 & \quad \forall j\in \items\\
& 0 \leq x_{ij} &\leq 1 & \quad \forall i\in \players,\forall j\in \items
\end{array}
\]

It is easy to see that, we can modify any given fractional solution $x$ to the Assignment-LP such
that for all $i\in\players$ it holds that $\sum_{j\in \items} x_{ij}p_{ij}\leq B_i$, without
decreasing its objective value; therefore, from now on we always assume that this is the case for any solution to the Assignment-LP we examine. Note that we will
require this only for our initial solution; at times we might want to violate this property.
Hence, let $\Val[x]=\sum\limits_{i,j} x_{ij}p_{ij}$, and for any $i\in \players$, let $\Val_i [x]=\sum\limits_{j\in\items} x_{ij} p_{ij}$.

For each player $i$, let $b_i=\sum\limits_{j\in \bigitems_i} x_{ij}$, $S_i=\sum\limits_{j\in\smallitems_i} x_{ij}p_{ij}$ and
$\alpha_i=\sum\limits_{j\in\items}x_{ij}p_{ij}/B_i$. In general, we will use either $x$ or $x$ with
a superscript to denote a solution; the same superscript will be applied to $b_i$, $S_i$ and $\alpha_i$ 
%to
%denote the corresponding value 
(e.g. for $x^1$ we will have $b^1_i$ and $S^1_i$). 

Given a fractional solution $x$ to the Assignment-LP, we will say the budget of player $i$ is
saturated if $\sum\limits_{j}x_{ij}p_{ij}\geq B_i$.

It is known (e.g. \cite{DBLP:journals/siamcomp/ChakrabartyG10}) that the integrality gap of the
Assignment LP is $3/4$, and we have matching polynomial time rounding algorithms; therefore, in order
to have a hope of beating this factor, we will need a stronger relaxation.

The second LP relaxation we will use is called the Configuration-LP. In this
relaxation, there is a variable $y_{i\conf}$ for every player $i\in\players$ and for every subset
of items $\conf\subseteq \items$; in an integer solution, $y_{i\conf}$ would be 1 if player $i$
received the set of items $\conf$, and 0 otherwise. The only constraints would be that every
player gets at most one set of items, and an item belongs to at most one assigned set of items. Let $w_i(\conf)=\min \{\sum\limits_{j\in \conf}
p_{ij},B_i\}$; the Configuration-LP can be formulated as follows:

\[
\begin{array}{rrl}
  \max & \sum_{i\in \players}\sum_{\conf\subseteq \items} w_i(\conf) y_{i\conf}&\\
\text{s.t.} & \sum_{\conf\subseteq \items} y_{i\conf}&\leq 1 \quad \forall i\in \players \\
& \sum_{i\in \players, \conf\subseteq \items: j\in \conf} y_{i\conf}&\leq 1 \quad \forall j\in
\items\\
& y_{i\conf}&\geq 0 \quad \forall i\in \players,\forall \conf\subseteq \items
\end{array}
\]

Of course, the Configuration-LP contains an exponential number of variables; however, it is
known \cite{DBLP:journals/siamcomp/ChakrabartyG10} that its fractional relaxation can be solved up
to any desired accuracy in polynomial time.

The best known lower bound on the integrality gap of the Configuration-LP is $3/4$, and
thus is not better than that of the Assignment-LP. However, it is
known \cite{DBLP:conf/ipco/KalaitzisMNPS14} that for certain restrictions of the MBA problem, the
integrality gap is larger than $3/4$, and hence this is an indication that the Configuration-LP might
actually be stronger than the Assignment-LP in general. This is the central question that we address.

Now, it is not hard to see (see, for example, \cite{DBLP:conf/ipco/KalaitzisMNPS14}) that we can transform any given solution to the
Configuration-LP
into a solution to the Assignment-LP of at least the same objective value (of course, the inverse is
not possible); hence, we will use $y$ to denote a solution to the Configuration-LP and
$x$ the corresponding solution to the Assignment-LP. We should note that throughout the paper,
our algorithms will only consider assigning $j$ to $i$ if $x_{ij}$ is in the support of $x$, i.e., if $x_{ij}>0$.

An important concept around which our approach will revolve, is that of canonical solutions: 
a solution $x$ to the Assignment-LP will be called  canonical if:
\begin{itemize}
\item[(a)] for all $i\in\players$ and all $j\in\bigitems_i$, $x_{ij}>0$ implies $p_{ij}=B_i$.
  \item[(b)] for all $i\in\players$, $\sum\limits_{j\in\bigitems_i} x_{ij}p_{ij}=\sum\limits_{j\in\smallitems_i} x_{ij}p_{ij}=B_i/2$.
\end{itemize}

\section{Overview and Organization}

Our next goal is to conduct an overview of our approach towards proving Theorem \ref{thm:main}. We start by 
giving a brief description of the main algorithm, in which we focus on its more important components.
A detailed description and analysis of the algorithm is given in Section \ref{sec:wrap}.
Afterwards, we explain what are the main novel ideas that were used in this algorithm, and what are the technical obstacles they overcome.

On the highest level, our main algorithm solves the Configuration-LP, projects the solution $y$ of the Configuration-LP to a solution $x$ of the Assignment-LP, 
and then performs a case analysis regarding the structure of fractional solution $x$. During this case analysis, all cases but the last one 
eventually lead to rounding $x$ using the Shmoys-Tardos (ST) algorithm; therefore it becomes clear that a strong analysis of this algorithm will be essential
in our approach.

Before discussing the main technical obstacles we will have to overcome while analyzing our main algorithm, let us briefly discuss the idea 
that led to its design. Previous work (\cite{DBLP:conf/ipco/KalaitzisMNPS14}) has provided an algorithm that projects Configuration-LP solutions to Assignment-LP solutions, and then
rounds these solutions to achieve a good approximation guarantee. However, this rounding requires these solutions to meet certain conditions; this is the 
reason we design an algorithm based on an extended case analysis, which will make sure these conditions are met. The problem is that, while some of these 
conditions can be met with standard techniques, it is more complicated to meet the following ones: that every item has a unique price for all the players it can 
be assigned to and that every item is either only assigned as big or only as small; meeting these conditions will
be the central technical problem we tackle.

Throughout the algorithm, the following theme comes up consistently: we take a fractional solution to the assignment LP, modify it, and then round it to
an integral one using the ST rounding algorithm. 
It is worth noting that these modifications will often force a player to exceed his budget. Since we are shifting items from one player to another, the first
major technical challenge we face is coming up with a suitable analysis of the ST algorithm. In particular, we want to have reasonably tight guarantees on
the expected value of the returned integral solution. This will give us a nice way of quantifying how much the performance of the ST algorithm changes for one 
player, even when slightly increasing or decreasing the fractional allocation of a single item to that player. 

In order to perform such an analysis, we come up with 
the concept of {\em worst-case arrangements of configurations}. Simply put, what we do is look at all the items fractionally assigned to a player, design
the absolutely worst probability distribution over the sets of items which are assigned to that player by the rounding algorithm, and then evaluate the 
player's expected payoff. While the pertaining analysis and guarantees are described in Section \ref{sec:STanalysis}, we note that 
the advantage of this approach is that it enables us to extract guarantees that depend on the structure of the fractional solution, i.e.,
the expected returned value of a player can be expressed as a function of how filled up his budget is, or how much of his LP value comes from big or small items.
Using these guarantees, and applying local modifications to the fractional solution, we are able to prove improved approximation guarantees when items do not have unique 
prices, which means that we can effectively enforce the condition of all items having one unique price.

The second main technical contribution of this paper addresses the problem of enforcing the condition that all items are either assigned only as big or only as small. Kalaitzis et al (\cite{DBLP:conf/ipco/KalaitzisMNPS14}) showed
how to achieve an improvement over the $3/4$-approximation factor when all items have unique prices and all players have the same budget, which is a special
case of the condition we want to enforce (always under the condition that items have unique prices). The problem that arises is that the techniques we use 
to enforce the 
unique prices condition can not be applied directly to this case. The reason is that, while in the previous case adding up the gain over many small local modifications
was sufficient to achieve our goal, in this case it is no longer possible; in fact, there are examples that show that applying small local modifications does not 
improve the performance of the ST algorithm at all, let alone provide an improvement that is provable using our guarantees. 

This is why we develop the concept of
{\em gain-amplifying players}. Specifically, what we do is enforce a certain structure on our fractional solution, such that local modifications now provide a gain that is 
sufficient for us to prove an improved approximation guarantee. In order to do this, we take advantage of the fact that, through our analysis of the ST algorithm,
we have developed performance guarantees that depend on how much of the LP value of a player comes from big and small items. Hence, the idea we employ is the following:
we will randomly partition the players into two sets, and then shift the allocation of items from one set of players to the other, according to whether these
items are assigned as big or small to the players in each set.
Regarding this modification, one can think of every player having 2 parameters, which define the expected value this player will receive from the ST
rounding algorithm. These parameters correspond to the LP value they receive from big and small items; then,
what we are trying to do is bring these parameters in a good range of values for a subset of the players, while not making them too bad for the rest.
This procedure will manage to change the structure of our fractional solution in such a way, 
that now we will be able to obtain a significant gain by applying a second round of local modifications. The above algorithm and its analysis appear
in Section \ref{sec:nup}.

Having overcome these two obstacles, we are able to complete our analysis of the algorithm behind Theorem \ref{thm:main}; each step of the algorithm is described and 
analyzed in detail in Section \ref{sec:wrap}.

\section{ST Rounding and Worst-Case Arrangements of Configurations}\label{sec:STanalysis}
As we have already explained, one of the most important parts of our approach is analyzing the ST rounding algorithm, 
and using this analysis to come up with strong approximation guarantees. In the following section, we introduce 
the ST rounding algorithm formally, and carry on with conducting 
its analysis, thus establishing the main technical lemmas we will use, 
concerning the performance of the ST rounding algorithm.

\subsection{The ST Rounding Algorithm}
Let us start with introducing the ST rounding algorithm. Given a fractional solution $x$ to the
Assignment LP, we create a bipartite graph $G=(U\cup V,E)$ as follows: first, $U$ consists of one node $u_j$ 
corresponding to every item $j$. Now, focus on one player $i$. We introduce $\lceil
\sum\limits_{j\in\items}x_{ij}\rceil$ nodes (also referred to as buckets) $v_{i1}...v_{ik}$ corresponding to $i$ in $V$. Next, go over
all the items $j$ such that $x_{ij}$ is non-zero, in order which is non-increasing with respect to
the price of $j$ for $i$. For each such item, connect it to the first copy of $i$, until the sum of
$x_{ij}$-s of connected items is at least 1; the first item for which this will happen will be
connected to the first and second copy of $i$. Now, continue in the same way for all items and
copies of $i$. 

After this procedure is over, observe that we have properly defined a bipartite graph $G$ and a
corresponding fractional solution to the bipartite matching problem, which lies inside the bipartite matching polytope $\{x\in\mathbb{R}^E:
\forall v\in U\cup V\ \sum\limits_{e:v\in e}x_e\leq 1, x\geq {\bf 0}\}$.

Hence,
since the corresponding
polytope is integral, we can decompose it into a convex combination of matchings, and sample one of
them. Finally, the items assigned to $i$ are those that are matched to its copies after the
sampling. One important property of this algorithm is that the probability that item $j$ is
assigned to $i$ is exactly $x_{ij}$.

From now on, let $\text{ST}[x]$ denote the expected value of the integral solution returned by the ST algorithm,
when it is given fractional solution $x$ as input; additionally, let $\text{ST}_i[x]$ denote the expected contribution
to the objective function corresponding to the items assigned to player $i$.

\subsection{ST Rounding Algorithm Analysis}
Next, we analyze the performance of the ST rounding algorithm; our approach is to
examine the sets a player receives due to the execution of the ST
algorithm, and determine what these sets look like when their expected value is
minimized.

First, let us introduce some definitions that will be used in the coming analysis. Let $x$ be a fractional
solution to the Assignment-LP. 
For a fixed player $i$, let $k=\lceil\sum\limits_{j\in\items} x_{ij}\rceil$ be the number of
buckets created by ST. A key observation is that
we can see the ST algorithm as picking a configuration at random from a set
of configurations which contain $k$ or
$k-1$ items in such a way that the marginals of $x$ are preserved, and such that every configuration
picks at most one
item per bucket. Let $\conf$ be the set of these configurations for
player $i$ (notice that this notation should not be confused with the one used
to describe the Configuration-LP), 
and let $V_C=\min\{B_i,\sum\limits_{j\in C}p_{ij}\}$ be the value of $C\in\conf$, i.e.,
the contribution of player $i$ to the objective value if he were to receive the items in $C$. 
Let $W_C=\sum\limits_{j\in C}p_{ij}-V_C$ be the unutilized value of $C\in\conf$, i.e., how much more than the 
budget of player $i$ the items in $C$ are valued by $i$. 
Let $z_{iC}$ be the probability of picking $C$, according to the ST rounding algorithm. 
In order to make a
worst case analysis, we have to argue about the way in which the total unutilized value will be
maximized. Already one can see that two of the most important quantities in this framework,
are $\E[V_\conf]$ and $\E[W_\conf]$, where the expectation is taken over the probability distribution
defined by $z$; the first quantity corresponds to the expected contribution of player $i$ to the
objective value, and the second one corresponds to how much value is unutilized by the ST algorithm on expectation.
Essentially, and this will become clearer later on, the second quantity measures how much value the rounding algorithm
loses.

\paragraph{Worst case arrangements of configurations}
Now, what we want to do, is find out how these configurations
look like in the worst case. 
We order $\conf$ in non-increasing order of $\sum\limits_{j\in C}p_{ij}$. 
By switching items among configurations, we can make a
series of changes that do not decrease the total unutilized value. This process goes as follows:
assuming all
configurations that are picked with non-zero probability are picked with the same probability (this
restriction comes with
no loss of generality if we consider some configurations appear more than once), let $C_l$ be the
item of $C\in\conf$
corresponding to the $l$-th bucket and let $C,C'\in\conf$ such that $p_{iC_l}>p_{iC'_l}$ and
$\sum\limits_{j\in
C}p_{ij}\leq \sum\limits_{j\in C'}p_{ij}$; then, switching $C_l$ with $C'_l$ will never decrease the
expected unutilized value.

Apply this operation iteratively until it can be applied no more; we will call the resulting $z_i$ and $\conf$ a worst-case arrangement
of configurations.
From the above discussion, the following facts become clear:
\begin{fact}\label{fact:confs}
  Any worst-case arrangement of configurations $(z_i,\conf)$ satisfies the marginals of the original Assignment-LP solution $x$;
furthermore, for all $l<l'$ we have $p_{iC_l}\geq p_{iC_{l'}}$, and for all $C,C'\in\conf$ and for all $t>0$, $\sum\limits_{j\in
C}p_{ij}\geq \sum\limits_{j\in C'}p_{ij}$ implies $p_{iC_t}\geq p_{iC'_t}$.
\end{fact}

\begin{fact}
 Given a worst-case arrangement of configurations $(z_i,\conf)$, and given any other arrangement of configurations $(z'_i,\conf')$ which
satisfies the marginals of the Assignment-LP solution $x$, we have 
$\sum\limits_{C\in\conf}z_{iC}V_C\leq\sum\limits_{C\in\conf'}z'_{iC}V_C$.
\end{fact}

% Observe that for all
% $C\in \conf_\bigitems$ and $C'\in\conf_\smallitems$, $\sum\limits_{j\in C}p_{ij}\geq
% \sum\limits_{j\in C'}p_{ij}$ (since the item of $C$ corresponding to the $l$-th bucket dominates the
% corresponding item
% of $C'$) and $\sum\limits_{j\in C\cap \smallitems}p_{ij}\leq \sum\limits_{j\in
% C'}p_{ij}$ (to see this, notice that by the design of the  worst-case arrangement of configurations,
% we know that for all $C,C'\in
% \conf$ and for all $l\in[k]$, we have $p_{iC_{l+1}}\leq p_{iC'_{l}}$).

\paragraph{Analysis}
Next, we will obtain our main approximation guarantees, which will be used in the majority of the
scenarios we will examine.

The main idea behind our analysis is the following fact: in a worst-case arrangement of configurations, 
the least valuable configuration will have value at least as much as the unutilized value of the configuration which maximizes unutilized value.
We build our analysis upon this observation, combined with the fact
that the expected unutilized value is the difference between the LP value and the expected returned value from the ST
algorithm. Therefore, in order to illustrate what is going to happen, one could say that most of our arguments are of the 
form: if the expected unutilized value is large (which means we lose a lot relative to the LP value), then the least valuable configuration
has large value, and therefore the total LP value must be larger than it actually is (a contradiction).

Let us begin with some definitions. Given a player $i$ and a worst-case arrangement of configurations $(z_i,\conf)$,
let $\conf_\bigitems$ be the set of configurations which contain a big item, and let $\conf_\smallitems=\conf\setminus\conf_\bigitems$. Next, let $\conf_W=\{C: \sum\limits_{j\in C}p_{ij}\geq B_i\}$.
Given the above definitions, let $v=\sum\limits_{C\in\conf_\smallitems\cap\conf_W} z_{iC}$ and let $w=\sum\limits_{C\in\conf_W} z_{iC}$ be
the probability of picking a configuration with non-negative unutilized value. We should point out that we can assume w.l.o.g. that
$b_i\leq 1$, since otherwise all our performance guarantees will hold anyway.

Now, let  $L=\sum\limits_{C\in\conf_W}z_{iC} W_{C}/w$ be the average
unutilized value over configurations with non-negative unutilized value,
let $L_\bigitems=\sum\limits_{C\in\conf_\bigitems\cap \conf_W}z_{iC} W_{C}/b_i$ be the average
unutilized value over big
configurations and let $L_\smallitems=\sum\limits_{C\in\conf_\smallitems\cap \conf_W} z_{iC} W_{C}/v$
be the average unutilized
value over small configurations with non-negative unutilized value.
Similarly, let
$G=\sum\limits_{C\in\conf\setminus\conf_W} z_{iC}
V_C/(1-w)$ be the average gain over configurations with zero unutilized value.

Given the above definitions, a very important observation, which will play an important role in the analysis to come, is 
that for any configurations $C,C'$ in a worst-case arrangement of configurations, $V_C\geq \min\{W_{C'},B_i\}$; 
this is due to the fact that for any two such configurations, the $l$-th largest item of $C$ is larger than the $l+1$-th largest item of
$C'$. This means that from now on, we can assume that for all configurations $C$ in a worst-case arrangement of configurations, $W_C\leq
B_i$; the contrary would imply that for any configuration $C'$ in the same arrangement of configurations we have that $V_C\geq B_i$, and
therefore the expected value player $i$ receives is $B_i$, i.e., player $i$ retrieves all of his LP value.

Now, the fact that for any configurations $C,C'$ in a worst-case arrangement of configurations, $V_C\geq W_{C'}$, implies 
$$
G\geq L_{\bigitems}
$$
while the definition of $L_{\bigitems}$, $L$ and $L_{\smallitems}$ along with Fact \ref{fact:confs} implies
$$
L_{\bigitems}\geq L\geq L_{\smallitems}
$$

We will now state the first core guarantee on the performance of ST rounding; let $\alpha=\frac{\sum\limits_{j\in\items} x_{ij}p_{ij}}{B_i}$:
\begin{lemma}\label{lem:ST}
 Given a fractional solution $x$ to the Assignment-LP, for all $i\in\players$ $\text{ST}_i[x]\geq (1-\alpha/4)\sum\limits_{j\in
\items}x_{ij}p_{ij}$.
\end{lemma}
The above lemma heavily depends on the following one, which we state independently because we intend to use it later on as well:
\begin{lemma}\label{lem:ST:intermediate}
 $\text{ST}_i[x]\geq B_i(\alpha-w(\alpha-w))$.
\end{lemma}

\begin{proof}
Fix a player $i$, and a worst-case arrangement of configurations $(z_i,\conf)$.
Since $\sum\limits_{j\in C}p_{ij}\geq B_i$ for all $C\in\conf_W$, we have that 
$$\text{ST}_i[x]\geq wB_i+(1-w)G$$
If $G=B_i$, the lemma directly follows; therefore, we will assume that $G<B_i$ from now on.

Observe that 
\begin{align}
 \alpha B_i=&\sum\limits_{C\in \conf}z_{iC}\sum\limits_{j\in C}p_{ij}\nonumber\\
=&\sum\limits_{C\in \conf_W}z_{iC}(B_i+W_C)+\sum\limits_{C\in \conf\setminus\conf_W}z_{iC}V_C\nonumber\\
=&w(B_i+L)+(1-w)G\nonumber
\end{align}
which implies
$$
wB_i +(1-w)G=\alpha B_i-wL
$$
Finally, observe that $L\leq (\alpha-w)B_i$; to see this, assume the contrary, i.e., assume that $L>(\alpha-w)B_i$. Then,
$\alpha B_i=w(B_i+L)+(1-w)G\geq w(B_i+L)+(1-w)L=wB_i+L>wB_i+(\alpha-w)B_i=\alpha B_i$, a contradiction. We have
\begin{align}
\text{ST}_i[x]\geq& wB_i+(1-w)G=\alpha B_i-wL \nonumber\\
\geq&\alpha B_i-w(\alpha-w)B_i=B_i(\alpha-w(\alpha-w))\nonumber
\end{align}
\end{proof}
To see how Lemma \ref{lem:ST:intermediate} implies Lemma \ref{lem:ST}, observe that basic calculus implies that 
$$
w(\alpha-w)\leq \alpha^2/4
$$
and hence we have that
$$
\text{ST}_i[x]\geq B_i\alpha(1-\alpha/4)
$$

Now, observing that $\alpha\geq 2$ implies that the configuration returned by the ST algorithm always has value at least  $B_i$ and that
$\alpha\in[1,2]$ implies $\alpha(1-\alpha/4)\geq 3/4$, a direct implication of Lemma \ref{lem:ST}, which implies that the ST algorithm
readily constitutes a $3/4$-approximation algorithm, is the
following:
\begin{lemma}\label{lem:ST3/4}
Given a fractional solution $x$ and a player $i$, $\alpha\leq 1$ implies $\text{ST}_i\geq \frac{3}{4} \Val_i [x]$, and $\alpha\geq 1$
implies $\text{ST}_i[x]\geq \frac{3}{4}B_i$.
\end{lemma}

From the above analysis, we easily get an improved performance guarantee when a significant amount of small items has a relatively large price. Specifically, consider the following assumption:
\begin{assumption:val-small}
 $$
 \sum\limits_{j:j\in\smallitems_i\wedge p_{ij}\geq B_i(1/2+\lambda)}x_{ij}\geq \epsilon
 $$
 for some small constants $\lambda,\epsilon>0$.
\end{assumption:val-small}
\noindent where $\lambda$ is a small constant which defines which items have a relatively large price even
though they are small for some player $i$, and $\epsilon$ is a constant determining whether the
contribution of such items to the LP value is significant or not.

We will prove that $i$ recovers more than a $3/4$-fraction of his LP-value:
\begin{lemma}\label{lem:ST2}
 Given a canonical solution $x$, if the Valuable Small Items Assumption holds for some player $i$, then 
$$
\text{ST}_i[x]\geq (3/4+\epsilon')\sum\limits_{j\in\items}x_{ij}p_{ij}
$$
for some $\epsilon'>0$.
\end{lemma}
\begin{proof}
 Due to $x$ being canonical, we know
that 
$$
\sum\limits_{j\in\smallitems} x_{ij}p_{ij}=B_i/2
$$ 
as well as that $\alpha=1$ and $w\geq 1/2$. Hence, Lemma \ref{lem:ST:intermediate} implies
$$
\text{ST}_i[x]\geq B_i(1-w(1-w))
$$
which is minimized at $w=1/2$. This means that if $w$ is far away from $1/2$, then the
lemma follows. Hence, from now on we assume $w\leq 1/2+\tau$, for some small constant $\tau\in[0,\epsilon]$.

Now, we know 
$$
\text{ST}_i[x]\geq wB_i+(1-w)G=B_i-wL
$$
which means that in order to prove the lemma, it suffices to prove that $wL<B_i/4$. 

Let $\epsilon'$ be a small positive constant; if $L\leq (1-w-\epsilon')B_i$, then basic calculus implies that $w(1-w)\leq 1/4$, and hence
$wL\leq w(1-w-\epsilon')B_i$ is strictly less than $B_i/4$.

On the other hand, if $L> (1-w-\epsilon')B_i$, then since $w\geq 1/2$, we can define a positive constant
$\lambda'=(1/2+\lambda-(1-w-\epsilon'))B_i$. Since $\tau\leq \epsilon$, we know that the configuration assigned to $i$ contains a small item
of value at least $(1-w+\lambda')B_i$ with probability at least $1/2+\epsilon-w$. Furthermore, we know $G\geq L$.
We have
\begin{align}
B_i=&\sum\limits_{i,j}x_{ij}p_{ij}=w(B_i+L)+(1-w)G\nonumber\\
\geq&w(B_i+L)+(1-w)L+(1/2+\epsilon-w)\epsilon'B_i\nonumber\\
>&B_i-\epsilon'B_i+(1/2+\epsilon-w)\lambda'B_i'\nonumber\\
\geq& B_i-\epsilon'B_i+(1/2+\epsilon-w)\lambda B_i\nonumber
\end{align}

Letting $\epsilon'$ become sufficiently small, and letting $\tau$ (and therefore $w$) become sufficiently small, the above expression is at
least $B_i$, which is a contradiction.
\end{proof}

Next, we will prove the last performance guarantee we will use. We start with the following lemma:
\begin{lemma}\label{lem:ST3}
  If $j\in\bigitems_i$ implies $p_{ij}=B_i$, and for all $j\in\smallitems_i$ $p_{ij}\leq B_i/2$, then  $\text{ST}_i[x]\geq b_iB_i+vB_i+(1-b_i-v)(S_i-vB_i/2)$.
\end{lemma}
\begin{proof}
Consider $C\in \conf_\bigitems$ which maximizes $W_C$ and $C'\in\conf_\smallitems$ which minimizes
$V_{C'}$, in the support of the worst-case arrangement $(z_i,\conf)$. 
If $W_C\geq B_i$, we have that
$V_{C'}=\min\{B_i,\sum\limits_{j\in C'}p_{ij}\}= B_i$, which implies $\text{ST}_i[x]=B_i$.

Otherwise, if $W_C<B_i$, we have that $V_{C'}=\min\{B_i,\sum\limits_{j\in C'}p_{ij}\}\geq W_C$, which implies
$G\geq L_{\bigitems}$.

Furthermore, since small items have price at most $B_i/2$ and for all
$C\in\conf_W\cap\conf_\smallitems,
C'\in\conf_\smallitems\setminus\conf_W$, the $l$-th largest item of $C'$ dominates the $l+1$-th
largest item of $C$,
we have $V_{C'}\geq B_i+W_{C}-B_i/2$. 
This fact implies $G\geq B_i+W_{C}-B_i/2$ for all $C\in \conf_\smallitems\cap \conf_W$, since $G$ is the expected value 
conditioned on picking a small configuration with zero unutilized value.
%, and any big configuration has value at least that of any small configuration. 
In turn, this implies $G\geq B_i+L_\smallitems-B_i/2=B_i/2+L_\smallitems$, since any big configuration has unutilized value at least that
of any small configuration. 
Now, since only small items
contribute to $L_\bigitems$ and $L_\smallitems$, the total fractional value of small items assigned
to $i$, i.e. $S_i$,
corresponds to the total unutilized value ($b_i L_\bigitems+vL_\smallitems$) plus the total returned
value from small configurations ($vB_i+(1-b_i-v)G$). Hence, since $S_i$ is the total value of small items,
$$
S_i=b_iL_\bigitems+v(B_i+L_\smallitems)+(1-b_i-v)G,
$$
and since $G\geq B_i+L_\smallitems-B_i/2$,
$$
S_i\leq b_iG+v(B_i/2+G)+(1-b_i-v)G.
$$
We conclude that
$$G\geq S_i-vB_i/2.$$

Finally, the value returned by ST is
\begin{align}
 \text{ST}_i[x]\geq& b_iB_i+vB_i+(1-b_i-v)G\nonumber\\
 \geq& b_iB_i+vB_i+(1-b_i-v)(S_i-vB_i/2)\nonumber
\end{align}
\end{proof}

From the above lemma, we get the following corollary:
\begin{corollary}\label{cor:ST3}
  If $j\in\bigitems_i$ implies $p_{ij}=B_i$, and for all $j\in\smallitems_i$ $p_{ij}\leq B_i/2$, then for $S_i/B_i\leq \frac{1+b_i}{2}$,
$\text{ST}_i[x]\geq b_iB_i+(1-b_i)S_i$. In particular, this is also
  true for $(S_i/B_i,b_i)\in[2/5,3/5]\times [2/5,1]$.
\end{corollary}
\begin{proof}
If $\conf_W\cap \conf_\smallitems =\emptyset$, we have that $v=0$ and the corollary follows directly
from Lemma
\ref{lem:ST}.

% since for all $C\in \conf_\bigitems$ and $C'\in\conf_\smallitems$,
% $\sum\limits_{j\in C\cap \smallitems}p_{ij}\leq \sum\limits_{j\in C'}p_{ij}$, we have that $v=0$
% and $G\geq
% L_\bigitems$; therefore, the total wasted value is maximized if $L_\bigitems=G=S_i$, an dw e have
% that
% $\text{ST}_i[x]\geq
% b_i+(1-b_i)S_i$.

Hence, let us assume $\conf_W\cap \conf_\smallitems \neq\emptyset$, and let $\text{ST}_i[x]\geq
b_iB_i+vB_i+(1-b_i-v)(S_i-vB_i/2)=f(v)$. In order to lower bound  $\text{ST}_i[x]$ we will place a
lower bound on
$f(v)$. We have
\[
\frac{df(v)}{dv}=vB_i+B_i\frac{1+b_i}{2}-S_i
\]

Since $f$ is a degree 2 polynomial in $v$, whose second degree coefficient is positive, it is minimized for a unique value of $v$; demanding
\[
\frac{df(v)}{dv}=vB_i+B_i\frac{1+b_i}{2}-S_i=0
\]
we get that $f$ is minimized at $v^*=S_i/B_i-\frac{1+b_i}{2}$. For $S_i\leq B_i\frac{1+b_i}{2}$, $v^*\leq
0$. Since  f is a degree 2 polynomial and $v^*\leq 0$, we get that $f$ is increasing in $[v^*,\infty]$.
Since the only valid values for $v$ are $[0,1]$,
 we get that when $S_i\leq B_i\frac{1+b_i}{2}$, $f$ is minimized
at $v=0$, in which
case we get
\[
f(0)=b_iB_i+(1-b_i)S_i
\]
In particular, for all $(S_i/B_i,b_i)\in[2/5,3/5]\times [2/5,1]$, we get that $S_i\leq B_i\frac{1+b_i}{2}$
and hence 
\[
\text{ST}_i[x]\geq f(0)=b_iB_i+(1-b_i)S_i
\]
\end{proof}

%\input ST
%\input gain
% \input ack
% \input nubp
% \input hbp
% \input nup

%\newpage

% \appendix
% \input ST
\section{Dealing with non-unique prices}\label{sec:nubp}
Next, we will try to show that the instances in which the prices of an item vary greatly guarantee a
better than $3/4$ performance for ST rounding on solutions which only satisfy mild technical assumptions. 
 Specifically, in the following we only assume that we are given a fractional solution $x$ to the Assignment-LP, for which the budget of every player is saturated; notice that this condition is satisfied by canonical solutions, whose importance we will see later on.

First of all, we have to define what we mean by saying that an item has prices that vary greatly. 
Intuitively, we would say that the prices of an item {\em do not} vary greatly, 
if there was a very small range of prices, such that, restricting the item to only be assigned to 
players whose valuation of the item is within that range, would not change the objective value of our fractional solution greatly; 
this will be the motivation behind our definition of unequally priced items.

More formally, let $x_j=\sum\limits_{i\in\players} x_{ij}$, and $w_j=\sum\limits_{i} x_{ij}p_{ij}/x_j$ 
be the average price of $j$. We will call an item $j$
$\mu$-unequally-priced if
$$
\mu\sum\limits_{i\in\players} x_{ij}p_{ij}\leq
\sum\limits_{i: p_{ij}\geq (1+\mu) w_j} x_{ij}p_{ij}
$$
where one should think of $\mu$ as being really close to 0. 
Intuitively, an item $j$ is $\mu$-unequally-priced if a substantial fraction of its
contribution to the LP value comes from its assignment to players that value it highly, instead of 
item $j$ being fractionally assigned to players that value it more or less the same.
Notice this definition only looks at pricing an item substantially higher than its average price;
however, after we show how we deal with this scenario, it will become clear that the symmetric case 
$$
\mu\sum\limits_{i\in\players} x_{ij}p_{ij}\leq
\sum\limits_{i: p_{ij}\leq (1-\mu) w_j} x_{ij}p_{ij}
$$
can be dealt analogously.

Let $N(\mu)$ be the set of
$\mu$-unequally-priced items.
Now, let us assume at least an
$\epsilon$-fraction of the LP value comes from the contribution of $\mu$-unequally-priced items:
\begin{assumption:unique-prices}
\[
\sum\limits_{j\in N(\mu)}\sum\limits_{i\in\players} x_{ij}p_{ij}\geq \epsilon
\sum\limits_{i,j} x_{ij}
p_{ij}
\]
\end{assumption:unique-prices}

% What we will do is the following: we already know that ST will return at least $3/4$-fraction of the LP value,
% therefore we will look at each non-uniformly-priced item individually, shift things around a bit, and show that the
% increase in the value that ST returns is a constant fraction of the original item's contribution, which will directly
% imply that summing these small increases we get a strictly better than $3/4$ guarantee. Notice that since we will only
% shift big items around, we only need to worry about the prices of these items, and not the total value of small items
% assigned on various machines. This is justified by the fact that the solution is canonical, 
% and therefore the small items always take up half the budget of
% the corresponding player. The fact that the guarantee from Corollary \ref{cor:ST} is linear in $b_i$ will help in the
% simplicity of the argument, since we will only have to care about the coefficients of the expression ($B_i-B_i/2$ for
% $b_i$, since our solution is canonical) and then add the individual differences.

Formally, the main result of this section is the following:
\begin{lemma}\label{lem:main-nubp}
 There is a polynomial time algorithm that, given a fractional solution $x$ satisfying the Non-Unique Prices Assumption and 
where for all players $i$ $\alpha_i=1$, returns a feasible solution
$x'$ such that 
$$
\text{ST}[x']\geq (3/4+c) \sum\limits_{i,j} x_{ij} p_{ij}
$$
for some $c>0$.
\end{lemma}

Putting the technicalities aside for the moment, the algorithm we will use, which we call Non-Unique Prices Algorithm,
is very straightforward. Looking at the definition of $\mu$-unequally-priced items, we know that if 
an item $j$ is $\mu$-unequally-priced, then there is a set of players which value $j$ very highly, 
and that the contribution to the LP value of the assignment of $j$ to these items is significant. 
Unavoidably, there is a similar set of players which value $j$ very low. Therefore, we will simply 
decrease the assignment of $j$ to the second set of players and increase its assignment to the 
first set of players. Repeating this procedure for all unequally priced items will output a 
modified fractional solution $x'$, for which the ST rounding algorithm will have a significantly improved performance.

Let us now delve more into the details of the above algorithm. Ultimately, our purpose is to 
apply Lemma \ref{lem:ST} to quantify the increase in performance we achieve. In order to be 
able to get some meaningful guarantee out of this lemma, we will need to preserve a certain 
invariant throughout the execution of the Non-Unique Prices Algorithm. Specifically, let $\alpha_i=\sum\limits_{j\in\items}x_{ij}p_{ij}/B_i$; 
from our technical assumptions, we know originally $\alpha_i=1$, for all players $i$. 
The sole invariant we will preserve throughout the execution of our algorithm, is that for any tentative solution $x^t$, 
and for any player $i$, $\alpha_i^t\in [1-\gamma, 1+\gamma]$, for some small constant $\gamma$ to be defined later on. 
The purpose behind preserving this invariant is the following: assume we are decreasing the assignment of some 
item $j$ to player $i$ by $\zeta$ and we are increasing its assignment to player $i'$ by $\zeta$. Then, 
applying Lemma \ref{lem:ST}, we know that the ST worst-case performance of player $i$ goes down by
$$
B_i \alpha^t_i(1-\alpha^t_i/4)-B_i(\alpha^t_i-\frac{\zeta p_{ij}}{B_i})(1-\frac{\alpha^t_i-\frac{\zeta p_{ij}}{B_i}}{4})
$$
$$
\leq p_{ij}(\zeta+\zeta^2/4-\zeta(1-\gamma)/2)
$$ 
where we used our invariant and the fact that $p_{ij}\leq B_i$. Similarly, 
the worst-case performance of player $i'$ goes up by
$$
B_{i'}(\alpha^t_{i'}+\frac{\zeta p_{i'j}}{B_{i'}})(1-\frac{\alpha^t_{i'}+\frac{\zeta p_{i'j}}{ B_{i'}
}}{4})-B_{i'}\alpha^t_{i'}(1-\frac{\alpha^t_{i'}}{4})$$
$$
\geq p_{i'j}(\zeta-\zeta(1+\gamma)/2-\zeta^2/4)
$$
where again we used our invariant and the fact that $p_{ij}\leq B_i$. Then 
$$p_{i'j}(\zeta-\zeta(1+\gamma)/2-\zeta^2/4)-p_{ij}(\zeta+\zeta^2/4-\zeta(1-\gamma)/2)$$ 
is a lower bound on the increase of the ST rounding performance due to this shifting; here we used the fact that the budgets of all players
are saturated, and therefore $\alpha_i=\alpha_{i'}=1$. It is 
clear that if $p_{i'j}\gg p_{ij}$, then our gain will be substantial, while if our invariant 
was not upheld, we would not be able to have the same guarantee. On the conceptual level, 
our description of the algorithm is complete; the rest of the section is devoted to choosing 
the right fractions to shift among players, such that in the above expression 
$\zeta^2\ll \zeta$, $1+\gamma\ll p_{i'j}/p_{ij}$ and still the total improvement in the ST rounding performance is significant. 

Let us now describe the Non-Unique Prices Algorithm formally; for all items $j\in N(\mu)$ do the following local update:
\begin{itemize}
\item Let $H_j=\{i\in\players: p_{ij}\geq (1+\mu)w_j\}$ and $L_j=\{i\in\players: p_{ij}\leq (1+\mu/2)w_j\}$. Let 
$h_j=\sum\limits_{i\in H_j}x_{ij}$ and let $l_j=\sum\limits_{i\in L_j}x_{ij}$.
\item For all $i\in L_j$, set $x'_{ij}\leftarrow x_{ij}(1-\frac{\mu(1+\mu)}{2+\mu}\frac{1}{10}\frac{l_j}{h_j})$.
\item For all $i\in H_j$, set $x'_{ij}\leftarrow x_{ij}(1+\frac{\mu(1+\mu)}{2+\mu}\frac{1}{10})$.
\end{itemize}

From the description of the algorithm and the fact that originally all the budgets are 
saturated, it follows that our invariant is upheld with $\gamma=\mu/10$ throughout 
the execution of Algorithm 1; to see this, observe that the Markov inequality and 
simple calculations imply that $h_j\leq 1/(1+\mu)$ and $l_j\geq \frac{\mu}{2+\mu}$. 
Furthermore, it is clear that $x'_j=x_j$ for all items $j$, and therefore $x'$ is a feasible solution to the Assignment-LP.

 Now, let $\text{ST}'_i[x]= B_i \alpha_i(1-\alpha_i/4)$ be the quantitative guarantee of 
 Lemma \ref{lem:ST} for player $i$, and  let $\text{ST}'[x]=\sum\limits_{i\in\players} \text{ST}'_i[x]$. 
 Finally, for $i\in L_j$ and $i'\in H_j$, let
 $$
 g(j,i,i')=
 p_{i'j}(\zeta(j,i,i')-\zeta(j,i,i')(1+\gamma)/2-\zeta^2(j,i,i')/4)$$$$-p_{ij}(\zeta(j,i,i')+\zeta^2(j,i,i')/4-\zeta(j,i,i')(1-\gamma)/2)
 $$
 be a lower bound on the gain in ST performance due to shifting item $j$ from $i$ to $i'$, where $\zeta(j,i,i')=x_{i'j}\frac{\mu(1+\mu)}{2+\mu}\frac{1}{10}$
is the fractional assignment of $j$ we shift from $i$ to $i'$. We have that 
 $$
 \text{ST}[x']\geq \text{ST}'[x']\geq \text{ST}'[x]+\sum\limits_{j\in N(\mu)}\sum\limits_{i\in L_j,i'\in H_j}g(j,i,i')
 $$
 Since from Lemma \ref{lem:ST3/4} we have that $\text{ST}'[x]\geq \frac{3}{4}\sum\limits_{i,j} x_{ij}p_{ij}$, 
 in order to prove Lemma \ref{lem:main-nubp}, it suffices to show that 
$$
\sum\limits_{j\in N(\mu)}\sum\limits_{i\in L_j,i'\in H_j}g(j,i,i')\geq c\sum\limits_{i,j} x_{ij}p_{ij}
$$
for some constant $c>0$. Hence, the proof of the following lemma completes the proof of Lemma \ref{lem:main-nubp}:
\begin{lemma}
$\sum\limits_{j\in N(\mu)}\sum\limits_{i\in L_j,i'\in H_j}g(j,i,i')\geq c\sum\limits_{i,j} x_{ij}p_{ij}$.
\end{lemma}
\begin{proof}
Fix $j\in N(\mu)$ and $i'\in H_j$; then $\sum\limits_{i\in L_j} g(j,i,i')$ is at least
\begin{align}
&p_{i'j}x_{i'j}\frac{\mu(1+\mu)}{(2+\mu)10}(1-(1+\mu/10)/2-(\frac{\mu(1+\mu)}{(2+\mu)10}\frac{1}{4})-\nonumber\\
&p_{ij}x_{i'j}\frac{\mu(1+\mu)}{(2+\mu)10}\frac{1}{10}(1+(\frac{\mu(1+\mu)}{(2+\mu)10})\frac{1}{4}-(1-\mu/10)/2)\nonumber
\end{align}
which is $\Omega(\mu)p_{i'j}x_{i'j}$ choosing $\mu$ small enough such that $\mu^2\ll \mu$, and since $p_{i'j}\geq \frac{1+\mu}{1+\mu/2}$. Summing over all $i'\in H_j$ we get 
$$
\sum\limits_{i\in L_j,i'\in H_j}g(j,i,i')=\Omega(\mu) \sum\limits_{i\in\players} x_{ij}p_{ij}
$$
since by definition of $N(\mu)$, $\sum\limits_{i\in H_j} x_{ij}p_{ij}\geq \mu \sum\limits_{i\in\players} x_{ij}p_{ij}$.
Finally, summing over all $j\in N(\mu)$ we get 
$$
\sum\limits_{j\in N(\mu)}\sum\limits_{i\in L_j,i'\in H_j} g(j,i,i')=\Omega(\mu,\epsilon) \sum\limits_{i,j} x_{ij}p_{ij}
$$
since from the Non-Unique Prices Assumption we know that $\sum\limits_{j\in N(\mu)} x_{ij}p_{ij}\geq \epsilon \sum\limits_{i,j}
x_{ij}p_{ij}$.
\end{proof}

\section{Dealing with Big-Small Items}\label{sec:nup}
From the previous section, it becomes clear that we can enforce the following restriction:
\begin{restriction:unique-prices}
All items $j\in\items$ have one unique price $p_j$ for all players they are fractionally assigned
to. 
\end{restriction:unique-prices}
Rounding fractional solutions to the class of instances of the MBA problem which satisfy the above restriction is one of the major
technical problems we have to overcome. Such instances capture much of the hardness of the general MBA problem (e.g., integrality gap
instances for both the Assignment-LP and the Configuration-LP, as well as instances that arise from NP-hardness reductions, fall into this
class), and solving such instances is a central point of our approach. 

Now, remember we can assume our fractional solution $x$ is canonical. Similarly, due to Lemma \ref{lem:ST2}, we can enforce the following
restriction:
\begin{restriction:val-small}
For all players $i$ and items $j\in\smallitems_i$, $p_{ij}\leq B_i/2$.
\end{restriction:val-small}

Finally, due to Lemma \ref{lem:ST} and the fact that in our original solution $x$, $\alpha_i=1$ for all $i\in\players$,
we can enforce the following restriction:
\begin{restriction:full-items}
For all items $j$, $\sum\limits_{i\in\players} x_{ij}\geq 9/10$.
\end{restriction:full-items}

In this section, we present the most crucial algorithm in this paper, the one that finally shows we can assume all items are
either big or small; the solutions that remain after we make this assumption we already know how to round.

Let $x$ be our updated fractional solution to the Assignment-LP, and let $x_j=\sum\limits_{i\in\players} x_{ij}$,
$x^\bigitems_j=\sum\limits_{i:j\in\bigitems_i} x_{ij}$ and $x^\smallitems_j=x_j-x^\bigitems_j$. We will call an item $j\in\items$
$\mu$-big-small if 
$$
\mu x_j\leq x_j^\bigitems \leq (1-\mu) x_j
$$
which clearly implies
$$
\mu x_j\leq x_j^\smallitems \leq (1-\mu) x_j
$$
where one should think of $\mu$ as being really close to 0. Let $M(\mu)$ be these
items. 
Intuitively, one should think of an item $j$ as being big-small if it is not exclusively assigned as a big 
or as a small item in $x$, even though its price is the same for any player it can be assigned to.

Now,
let us assume that at least an $\epsilon$-fraction of the LP value comes from such
items:
\begin{assumption:uniform-prices}\label{assumption:nup}
$$
\sum\limits_{j\in M(\mu)}\sum\limits_{i\in \players} x_{ij} p_j \geq \epsilon \sum\limits_{i,j} x_{ij} p_{j}
$$ 
\end{assumption:uniform-prices}

The main result of this section is:
\begin{lemma}\label{lem:main-nup}
 There is a polynomial time algorithm that, given a canonical solution $x$ satisfying the Big-Small Items Assumption and the above
restrictions, returns a feasible solution $x'$ such that
$$
\text{ST}[x']\geq (3/4+\gamma)\sum\limits_{i,j} x_{ij}p_{ij}
$$
for some $\gamma>0$.
\end{lemma}

Now, the big question is, what does an algorithm satisfying Lemma \ref{lem:main-nup} look like? First, we consider the available tools; the performance guarantee on the ST algorithm we want to use is that of Corollary \ref{cor:ST3}:
$$
\text{ST}_i[x]\geq b_iB_i+(1-b_i)S_i
$$
The first idea that comes to mind, is to shift the assignment of some items from some players to others, and use this, along with the fact that originally $b_i=S_i/B_i=1/2$ for all players $i$, to guarantee some improvement in the ST performance {\em for every such item}. While this idea is in the right direction, such a simple approach will not work; for example, if we shift the assignment of just a single item, then we will not be able to place a good lower bound in the ST performance. In fact, there are cases in which the ST performance will actually {\em not increase}.

In order to overcome this obstacle, we will design an algorithm which consists of two phases, a preprocessing phase and a main phase. From the design of these phases, we will guarantee that the following invariant is upheld throughout the execution of the algorithm, for any tentative solution $x^t$:
\begin{itemize}
\item For all players $i$, the conditions of Corollary \ref{cor:ST3} are satisfied.
\item For all players $i$, $S^t_i\leq B_i/2$.
\item There exists a set of players $G$ (green) and a set of players $R$ (red), such that $G\cup R=\players$, $\forall i\in G$ $b^t_i\leq 1/2$ and $\forall i\in R$ $b^t_i\geq 1/2$.
\item $\text{ST'}[x^t]\geq \text{ST'}[x]$, where $\text{ST'}_i[x]=b_i B_i+(1-b_i)S_i$ and $\text{ST'}[x]=\sum\limits_{i\in\players} \text{ST'}_i[x]$.
\end{itemize}

Having ensured the above, the question now becomes, how do we prove an improvement in the ST performance? The answer to this question lies within the design of our preprocessing phase. Specifically, the preprocessing phase will guarantee that there exist players in $R$ for which $b_i$ will be at least $\frac{1+c}{2}$, for some constant $c>0$ (in fact, this will hold in expectation for all players in $R$). Then, we will leverage the existence of items in $M(\mu)$ in the main phase: we will shift the assignment of items in $M(\mu)$ from players in $R$ in which these items are small to players in $R$ in which these items are big. The invariant of our algorithm will then directly imply, in conjuction with Corollary \ref{cor:ST3}, that the performance of the ST rounding algorithm is strictly improved.

Let us now proceed with describing the main algorithm of this section, which we call Big-Small Items Algorithm, formally. First,  we
describe the preprocessing phase of the algorithm: in this phase, a random partition of the players is chosen, and according to this
partition we apply certain local updates to our fractional solution:
\begin{itemize}
\item Choose a partition of the players into sets $R$ and $G$ uniformly at random.
\item For all $j\in\items$, and for all $i\in G:j\in\bigitems_i$ and $i'\in R:j\in\bigitems_{i'}$, set $x'_{ij}=
x_{ij}-\frac{x_{ij}}{100}\frac{x_{i'j}}{x_j-x_{ij}}$ and 
$x'_{i'j}= x_{i'j}+\frac{x_{ij}}{100}\frac{x_{i'j}}{x_j-x_{ij}}$.
\item  For all $j\in\items$, and for all $i\in R:j\in\smallitems_i$ and $i'\in R:j\in\bigitems_{i'}$, set 
$x'_{ij}=
x_{ij}-\min\{\frac{x_{ij}}{100}\frac{x_{i'j}}{x^\bigitems_j},\frac{x_{ij}}{100}\frac{x_{i'j}}{
x^\smallitems_j}\}$ and
$x'_{i'j}=
x_{i'j}+\min\{\frac{x_{ij}}{100}\frac{x_{i'j}}{x^\bigitems_j},\frac{x_{ij}}{100}\frac{x_{i'j}}{
x^\smallitems_j}\}$.
\end{itemize}

Due to the numbers chosen in the design of the preprocessing phase, the algorithm's invariant is upheld (observe that due to the Fully
Assigned Items Restriction, $x_j\geq 9/10$ and $\max\{x_{j}^\bigitems,x_j^\smallitems\}\geq 9/20$, and due to $x$ being canonical,
$x_{ij}\leq 1/2$), and Corollary \ref{cor:ST3} implies the following lemma:
\begin{lemma}\label{lem:nup:phase1}
 For any $x^1$ that might be the result of the preprocessing phase, $\text{ST'}[x^1]\geq
\text{ST'}[x]$.
\end{lemma}

Therefore,  we lose nothing from the preprocessing phase in terms of performance. However, we also cannot guarantee that we gain something
in terms of performance.
However, we will see that we gain
something in terms of structure of the solution, which we will use later on.

Let $x^1$ be the tentative solution after the preprocessing phase; we start off with the following lemma:
\begin{lemma}\label{lem:nup:Delta:independence}
Consider $j\in\items$, $i$ such
that $j\in\smallitems_i$ and $i'$ such that $j\in\bigitems_{i'}$;
 then
 $$\E\left[(1-b^1_i)\middle| i,i'\in R\right]=\E\left[(1-b^1_i)\middle| i\in R\right]$$
\end{lemma}
\begin{proof}
 Since by the Unique Prices Restriction, items only have one price for any player they
can be assigned to, if $j\in\smallitems_i$ and $j\in\bigitems_{i'}$, then $x$ being canonical and the Cheap Small Items Restriction  imply that $B_i>B_{i'}$.
However, $b^1_i$ will only depend on the random choices of the Big-Small Items Algorithm for players
$i''$ such that $B_{i''}=B_i$ (i.e., some item is big both for $i$ and  $i''$) or for players 
$i''$ such that $B_{i''}>B_i$ (i.e., some item is big for $i$ and small for $i''$). Therefore,
$b^1_i$ will not depend on the random choices of the algorithm for $i'$, and the claim
follows.
\end{proof}

The most crucial lemma of our whole approach is the following:
\begin{lemma}\label{lem:nup:bi}
For all $i\in\players$, $\E\left[b^1_i\middle| i\in R\right] \geq \frac{1+c}{2}$, for some constant $c>0$.
\end{lemma}
\begin{proof}
Since $x$ is originally canonical, we know that originally $b_i=1/2$, and by the invariants of
the Big-Small Items Algorithm, we know that $b_i$ can only increase throughout the execution of the algorithm when $i$ is red.
Therefore, in order to prove the desired claim, it suffices to prove that 
\begin{align}
 \forall j\in\bigitems_i\ \E\left[x^1_{ij}\middle| i\in R\right]\geq (1+c)x_{ij}\label{eqn:lem:nup:bi:claim}
\end{align}

Let  $I(A)$ be the indicator variable of event $A$. We have
\begin{align}
&\E\left[\sum\limits_{i'\neq i:j\in\bigitems_{i'}}I(i'\in
G)\frac{x_{i'j}}{100}\frac{x_{ij}}{x_j-x_{ i'j } } \middle| i\in R\right]\nonumber\\
=& \sum\limits_{i'\neq i:j\in\bigitems_{i'}}\E\left[I(i'\in
G)\middle| i\in R\right]\frac{x_{i'j}}{100}\frac{x_{ij}}{x_j-x_{ i'j } } \nonumber\\
=&\sum\limits_{i'\neq i:j\in\bigitems_{i'}}\E\left[I(i'\in
G)\right]\frac{x_{i'j}}{100}\frac{x_{ij}}{x_j-x_{ i'j } }\nonumber\\
=&\sum\limits_{i'\neq i:j\in\bigitems_{i'}}\frac{x_{i'j}}{200}\frac{x_{ij}}{x_j-x_{ i'j }
}\nonumber\\
\geq&\sum\limits_{i'\neq i:j\in\bigitems_{i'}}\frac{x_{i'j}}{200}\frac{x_{ij}}{x_j}
\nonumber\\
=&\frac{x_{ij}}{200}\frac{x_j^\bigitems-x_{ij}}{x_j}&\nonumber
\end{align}
due to the fact that $i\in R$ and $i'\in G$ are independent events that happen with probability $1/2$.
On the other hand, we have
\begin{align}
 &\E\left[\sum\limits_{i'\neq i:j\in \smallitems_{i'}} I(i'\in R)
\frac{x_{ij}}{100}\min\{\frac{x_{i'j}}{x^\bigitems_j},\frac{ x_{ i'j } } {
x^\smallitems_j}\}\middle| i\in R\right]
\nonumber\\
=&\sum\limits_{i'\neq i:j\in \smallitems_{i'}} \E\left[I(i'\in R)\middle| i\in R\right]
\frac{x_{ij}}{100}\min\{\frac{x_{i'j}}{x^\bigitems_j},\frac{ x_{ i'j } } {
x^\smallitems_j}\}\nonumber\\
=&  \sum\limits_{i'\neq i:j\in \smallitems_{i'}} 
\min\{\frac{x_{ij}}{200}\frac{x_{i'j}}{x^\bigitems_j},\frac{x_{ij}}{200}\frac{ x_{ i'j } } {
x^\smallitems_j}\}\nonumber\\
 \geq&\sum\limits_{i'\neq i:j\in \smallitems_{i'}} 
\frac{x_{ij}}{200}x_{i'j}\nonumber\\
=& \frac{x_{ij}}{200}x_j^\smallitems&\nonumber
\end{align}
since $\Pr[i'\in R]=1/2$.

We have that $\E\left[x^1_{ij}-x_{ij}\middle| i\in R\right]$ is at least
$$
\E\left[\sum\limits_{i'\neq i:j\in\bigitems_{i'}}I(i'\in
G)\frac{x_{i'j}}{100}\frac{x_{ij}}{x_j-x_{ i'j } } \middle| i\in R\right]+$$
$$
 \E\left[\sum\limits_{i'\neq i:j\in \smallitems_{i'}} I(i'\in R)
\min\{\frac{x_{ij}}{100}\frac{x_{i'j}}{x^\bigitems_j},\frac{x_{ij}}{100}\frac{ x_{ i'j } } {
x^\smallitems_j}\}\middle| i\in R\right]
$$
which is at least 
$$
\frac{x_{ij}}{200}\frac{x_j^\bigitems-x_{ij}}{x_j}+ \frac{x_{ij}}{200}x_j^\smallitems
$$
Due to the Fully Assigned Items Restriction, we know that $\max\{x^\bigitems_j,x_j^\smallitems\} \geq 9/20$, and due to $x$ being
canonical, we know that $x_{ij}\leq 1/2$; therefore, (\ref{eqn:lem:nup:bi:claim}) follows, and the lemma is proved.
\end{proof}
The above lemma is crucial for the following reason: the existence of players for which $b_i$ is strictly larger than $1/2$, implies that
decreasing the assignment of small items to those players will have a smaller impact on the performance of the ST rounding algorithm.
Therefore, one could view these players as {\em gain-amplifying}, since we are able to lose less when we remove small items from these
players, than what we gain when we reassign these items as big to other players.

We are now ready to describe the main phase of our algorithm; let $x$ be our updated fractional solution:
\begin{itemize}
\item For all $j\in M(\mu)$, and for all $i\in R:j\in\smallitems_i$ and $i'\in R:j\in\bigitems_{i'}$, set 
$x'_{ij}=
x_{ij}-\min\{\frac{x_{ij}}{100}\frac{x_{i'j}}{x^\bigitems_j},\frac{x_{ij}}{100}\frac{x_{i'j}}{
x^\smallitems_j}\}$ and
$x'_{i'j}=
x_{i'j}+\min\{\frac{x_{ij}}{100}\frac{x_{i'j}}{x^\bigitems_j},\frac{x_{ij}}{100}\frac{x_{i'j}}{
x^\smallitems_j}\}$.
\end{itemize}

Finally, let us analyze the gain and loss from the main phase. We will do so by analyzing the
gain or loss from a single modification by the main phase, and adding up. Consider $j\in M(\mu)$, $i$ such
that $j\in\smallitems_i$ and $i'$ such that $j\in\bigitems_{i'}$. 
The main phase of the Big-Small Items Algorithm applies two modifications corresponding to $j$ and $i,i'$: it decreases
$x_{ij}$ and increases $x_{i'j}$.
Let $x^t$ be the temporary
solution kept by the Big-Small Items Algorithm before the first modification (i.e., the decrease of $x_{ij}$) in the main phase that
corresponds to
$j$, $i$ and $i'$, and let $x^{t'}$ be the one after the modification. Then, we define
$$
\Deltah^-_{i,i'}(j)=\text{ST'}[x^{t'}]-\text{ST'}[x^t]
$$
to be the difference in ST performance due to this modification. Since according to the invariant of our algorithm, $b_i$ can only increase
for $i\in R$, according to Corollary \ref{cor:ST3} we have 
\begin{align}
\Deltah^-_{i,i'}(j)=&I(i,i'\in R)(x'_{ij}-x_{ij})p_j(1-b_i)\nonumber\\
\geq&-I(i,i'\in R)p_j\min\{\frac{x_{ij}}{100}\frac{x_{i'j}}{x^\bigitems_j},\frac{x_{ij}}{100}\frac{x_{ i'j }
}{x^\smallitems_j}\}/2\nonumber
\end{align}

Similarly,  let $x^t$ be the
temporary solution kept by the Big-Small Items Algorithm before the second modification (i.e., the increase of $x_{i'j}$) of the main
phase that
corresponds to $j$, $i$ and $i'$, and let $x^{t'}$ be the one after the modification. Then, we
define
$$
\Deltah^+_{i,i'}(j)=\text{ST'}[x^{t'}]-\text{ST'}[x^t]
$$
to be the difference in ST performance due to this modification.  Since according to the invariant of our algorithm, $S_{i'}$ can only
decrease, according to Corollary \ref{cor:ST3} we have 
\begin{align}
\Deltah^+_{i,i'}(j)=&I(i,i'\in R)(x'_{i'j}-x_{i'j})p_j(1-S_{i'}/B_{i'}) \nonumber\\
\geq&I(i,i'\in R)\min\{\frac{x_{ij}}{100}\frac{x_{i'j}}{x^\bigitems_j},\frac{x_{ij}}{100}\frac{x_{i'j}}{
x^\smallitems_j}\}p_j/2\nonumber
\end{align}

Notice that
$$
\text{ST'}[x^2]-\text{ST'}[x^1]=\sum\limits_{j\in\items}\sum\limits_{i:j\in\smallitems_i}\sum\limits_{
i':j\in\bigitems_{i'}} \Deltah^+_{i,i'}(j)+\Deltah^-_{i,i'}(j)
$$
where $x^1$ and $x^2$ are the tentative solutions after the preprocessing and main phases respectively.

Since $i$ and $i'$ are placed into $R$ independently and with probability $1/2$, and since
$$
\Deltah^-_{i,i'}(j)=-I(i,i'\in R)\min\{\frac{x_{ij}}{100}\frac{x_{i'j}}{x^\bigitems_j},\frac{x_{ij}}{100}\frac{x_{i'j}}{
x^\smallitems_j}\}p_j(1-b_i)
$$
$$
\Deltah^+_{i,i'}(j)\geq I(i,i'\in
R)\min\{\frac{x_{ij}}{100}\frac{x_{i'j}}{x^\bigitems_j},\frac{x_{ij}}{100}\frac{x_{i'j}}{
x^\smallitems_j}\}p_j/2
$$
we have the
following corollary:
\begin{corollary}\label{cor:nup:Delta}
For $x^t$ as specified in the definition of $\Deltah^-_{i,i'}(j)$,
 \begin{align}
 &\E\left[\Deltah^-_{i,i'}(j)\right] \nonumber\\
 \geq&- \E\left[(1-b^t_i)\middle| i,i'\in R\right] p_j
\min\{\frac{x_{ij}}{100}\frac{x_{i'j}}{x^\bigitems_j},\frac{x_{ij}}{100}\frac {x_{ i'j }
}{x^\smallitems_j}\}/2\nonumber
 \end{align}
 Furthermore,
 $$
 \E\left[\Deltah^+_{i,i'}(j)\right]\geq
\min\{\frac{x_{ij}}{100}\frac{x_{i'j}}{x^\bigitems_j},\frac{x_{ij}}{100}\frac{x_{i'j}}{
x^\smallitems_j}\}p_j/4.
 $$
\end{corollary}
Notice the dependence on the term $\E\left[(1-b^t_i)\middle| i,i'\in R\right]$; combined with Lemma \ref{lem:nup:bi}, this dependence will
enable
us to place a good guarantee on the performance increase of the ST rounding algorithm.
Specifically, from Corollary \ref{cor:nup:Delta}, Lemma \ref{lem:nup:Delta:independence} and Lemma
\ref{lem:nup:bi} we get the following Corollary:
\begin{corollary}\label{cor:nup:gain}
 $$
 \E\left[\Deltah^-_{i,i'}(j)+\Deltah^+_{i,i'}(j)\right]\geq c
\min\{\frac{x_{ij}}{100}\frac{x_{i'j}}{x^\bigitems_j},\frac{x_{ij}}{100}\frac{x_{i'j}}{
x^\smallitems_j}\}p_j/4.
 $$ 
\end{corollary}

The above corollary implies that we gain a significant performance improvement for every item $j$ for which
$\min\{\frac{x_{ij}}{100}\frac{x_{i'j}}{x^\bigitems_j},\frac{x_{ij}}{100}\frac{x_{i'j}}{
x^\smallitems_j}\}$ is large enough. The next lemma exploits the fact that items in $M(\mu)$ satisfy this condition, in order to prove that
for every item in $M(\mu)$, we are able to guarantee an improvement in performance which is significant (compared to that item's
contribution to the LP value):
\begin{lemma}\label{lem:nup:Na}
 For $j\in M(\mu)$,
 $$
 \sum\limits_{i:j\in
\smallitems_i}\sum\limits_{i':j\in\bigitems_{i'}}\E\left[\Deltah^-_{i,i'}(j)+\Deltah^+_{i,i'}
(j)\right]\geq \frac{9c\mu^2}{4000}\sum\limits_{i\in\players} x_{ij}p_j.
 $$
\end{lemma}
\begin{proof}
 We have
 \begin{align}
  &\sum\limits_{i:j\in
\smallitems_i}\sum\limits_{i':j\in\bigitems_{i'}}\E\left[\Deltah^-_{i,i'}(j)+\Deltah^+_{i,i'}
(j)\right] \nonumber\\
\geq&\sum\limits_{i:j\in
\smallitems_i}\sum\limits_{i':j\in\bigitems_{i'}}
c \min\{\frac{x_{ij}}{100}\frac{x_{i'j}}{x^\bigitems_j},\frac{x_{ij}}{100}\frac{x_{i'j}}{
x^\smallitems_j}\}p_j/4&\nonumber\\
 \geq& \sum\limits_{i:j\in
\smallitems_i}\sum\limits_{i':j\in\bigitems_{i'}}
c\frac{x_{ij}x_{i'j}}{400}p_j&\nonumber\\
=&\frac{cp_j}{400}
\sum\limits_{i:j\in
\smallitems_i}x_{ij}\sum\limits_{i':j\in\bigitems_{i'}}x_{i'j}&\nonumber\\
=&\frac{cp_j}{400} x_j^\bigitems x_j^\smallitems\nonumber\\
 \geq&\frac{cp_j}{400} (\mu x_j)^2\nonumber \\
\geq& \frac{cp_j}{400} \frac{9\mu^2 }{10} x_j\nonumber\\
=& \frac{9c\mu^2}{4000}\sum\limits_{i\in\players} x_{ij}p_j\nonumber
 \end{align}
 Here, the first inequality holds 
due to Corollary \ref{cor:nup:gain}, the second inequality holds since 
$x_j^\bigitems\leq 1$ and $x_j^\smallitems\leq 1$, the fifth inequality holds 
due to the definition of items in $M(\mu)$, and the sixth inequality holds 
due to the Fully Assigned Items Restriction.

\end{proof}

Summing up $$
 \sum\limits_{i:j\in
\smallitems_i}\sum\limits_{i':j\in\bigitems_{i'}}\E\left[\Deltah^-_{i,i'}(j)+\Deltah^+_{i,i'}
(j)\right]\geq
\frac{9c\mu^2}{4000}\sum\limits_{i\in\players} x_{ij}p_j$$ 
over all items $j\in M(\mu)$, we get that 
$$
\text{ST}'[x^2]\geq \text{ST}'[x]+\frac{9\epsilon c\mu^2}{4000}\sum\limits_{i,j} x_{ij}p_{ij}
$$
and since $\text{ST}'[x]\geq \frac{3}{4}\sum\limits_{i,j}x_{ij}p_{ij}$ (due to Lemma \ref{lem:ST3/4}), Lemma
\ref{lem:main-nup} follows.

%%%%%%%%%%%%%%%%%%%%%%%%%%%%%%%%%%%%%%%%%%%%%%%%%

%\input wrap
% \input nubp-app
% \input hbp-app
% \input nup-app
% \input theoremproof
\section{Putting everything together}\label{sec:wrap}

Our results so far tell us intuitively why Theorem \ref{thm:main} holds. In this last section, we describe the algorithm behind proving Theorem \ref{thm:main},
and put together all the technical facts we proved so far in order to analyze this algorithm.

The first step of the main algorithm is finding a solution $y$ to the Configuration-LP, and projecting $y$ to a solution $x$ of the Assignment-LP as
done in \cite{DBLP:conf/ipco/KalaitzisMNPS14}.
As a result of how this projection is performed, $x$ has the following property: for any player $i$, and any item $j\in \smallitems_i$,
$x_{ij}+\sum\limits_{j'\in \bigitems_i}x_{ij'}\leq 1$.
 Then, the algorithm performs an extensive case analysis regarding the structure of fractional solution $x$. 
Since every part of the algorithm is analyzed formally and in detail later on, we choose to be a bit informal at this point, refraining 
from giving the formal definitions, but instead giving intuitive ones (e.g., we might say "many agents do not have a full budget" instead of saying "an $a$-fraction
of the LP value comes from players whose budget is filled up to a $b$-fraction, for some constants $a,b$"):
\begin{enumerate}
 \item (See Section \ref{sec:wrap:full}) If many agents do not have a full budget, then run the ST rounding algorithm; otherwise, modify fractional solution $x$ by removing players that do not have a full budget. 
 \item (See Section \ref{sec:wrap:unique}) If many items are fractionally assigned to players with significantly different valuations for these items, then
 modify fractional solution $x$ appropriately (see Section \ref{sec:nubp}) and run the ST rounding algorithm; otherwise, remove such items from fractional solution $x$.
 \item (See Section \ref{sec:wrap:canonical}) If $x$ is far from being a canonical solution, run the ST rounding algorithm. Otherwise, we remove the agents that collide with
 the canonical solution definition, and trim $x$ into being canonical.
 \item (See Section \ref{sec:wrap:valuable-small}) If many items are assigned as small with a relatively high price (i.e., much larger than half of the budget),
 run the ST rounding algorithm; otherwise, remove such items from $x$.
 \item (See Section \ref{sec:wrap:fully-assigned}) If many items are fractionally not fully assigned, increase their fractional allocation (since at this
 point agents have full budgets, some agents will go over their budgets), and run the ST rounding algorithm; otherwise, remove such items from fractional solution $x$.
 \item (See Section \ref{sec:wrap:non-unique}) If many items are assigned both as big and as small, modify fractional solution $x$ according to
 Section \ref{sec:nup}, and then run the ST rounding algorithm; otherwise, remove all such items from $x$.
 \item (See Section \ref{sec:wrap:main}) Since all previous conditions are not satisfied, run the negatively correlated rounding algorithm from \cite{DBLP:conf/ipco/KalaitzisMNPS14}.
\end{enumerate}

% \sout{
% Our results so far tell us intuitively why Theorem \ref{thm:main} holds. For completeness, in the next section we will proceed to show how
% the constants that have appeared in our analysis so far should be picked, and in which order, and put all of our technical facts together to
% prove Theorem \ref{thm:main}. }

Let $\Opt$ be the LP value of
our original Assignment-LP solution $x$; in the following we will overload notation, in the sense that $x$ will always refer to the most
updated fractional solution, while $\Opt$ will always refer to the LP value of the original, since this is the upper bound towards which we
are comparing our algorithms against. Finally, in the following section, we might sometimes refer to inserting fake items into some LP
solution; this means that we will create an imaginary item and assign it to some player, in order to ensure the structure of our LP
solution looks like we want it to look. These items will participate in the execution of the rounding algorithm, but of course will be
removed before we return the final integral solution. We say that an item that is not fake is real.
\subsection{Dealing with not full budgets}\label{sec:wrap:full}
First, let us start with ensuring the following condition: for all
$i\in\players$,
$\sum\limits_{j\in\items}x_{ij}p_{ij}=B_i$. 
The technical lemma we will use to guarantee this condition is the following:
\begin{lemma}\label{lem:conditiona}
Let $\epsilon,\epsilon_1>0$ be small constants and $x$ be a fractional solution to the Assignment-LP. If at least an $\epsilon_1$ 
fraction of the LP value of $x$ comes from players $i$ such that  $\sum\limits_{j\in\items} x_{ij}p_{ij} \leq (1-\epsilon)B_i$, 
then $\text{ST}[x]\geq (3/4+c)\Val[x]$, for some constant $c>0$.
\end{lemma}

The above lemma follows directly from the following lemma from \cite{DBLP:conf/ipco/KalaitzisMNPS14} (in the paper, it appears as Lemma 3):

\begin{lemma}
\label{lem:not-fully-assigned}
  Let $\epsilon>0$ be a small constant and consider player $i$ such that
  $\sum\limits_{j\in\items} x_{ij}p_{ij} \leq (1-\epsilon)B_i$. Then $\text{ST}_i [x] \ge \frac{3+\epsilon/5}{4}
  \sum\limits_{j\in\items} x_{ij}p_{ij}$.
\end{lemma}

Therefore, if the requirements set by Lemma \ref{lem:conditiona} are satisfied, we can run the ST rounding algorithm on $x$ and 
the returned integral solution will have expected value at least $(3/4+c)\Opt$, for some constant $c>0$; 
in this case we would be done. 

Otherwise, we will remove all players $i$ for which 
$\sum\limits_{j\in\items} x_{ij}p_{ij} \leq (1-\epsilon)B_i$ (losing an $\epsilon_1$ fraction of the LP value). 
Then, we will lower the budget of every remaining player, in order to guarantee that for every remaining player $i$, 
$\sum\limits_{j\in\items} x_{ij}p_{ij} =B_i$; the LP value of $x$ is not affected by this modification.

 In total, all we lose from this step is an $\epsilon_1$ fraction of the LP value of $x$, i.e., 
 $$
 \sum\limits_{i,j}x'_{ij}p_{ij}\geq (1-\epsilon_1) \Opt
 $$

 \subsection{Dealing with Non-Unique Prices}\label{sec:wrap:unique}
 Next, since we have ensured that for all players $i$ $\sum\limits_{j\in\items} x_{ij}p_{ij}=B_i$, we are ready to put to use the analysis
behind Lemma \ref{lem:main-nubp}. Specifically, let $x$ be our updated fractional solution, and let us assume that the Non-Unique Prices
Assumption is true, i.e., that

\[
\sum\limits_{j\in N(\mu)}\sum\limits_{i\in\players} x_{ij}p_{ij}\geq \epsilon_2
\sum\limits_{i,j} x_{ij}
p_{ij}
\]
for some small constants $\mu,\epsilon_2>0$.

Then, according to Lemma \ref{lem:main-nubp}, there is an algorithm which returns an integral solution of
total value at least $(3/4+\Omega(\mu,\epsilon_2))\sum\limits_{i,j} x_{ij}p_{ij}$; by carefully selecting $\mu$ and $\epsilon_2$ to be much
larger than $\epsilon_1$, this implies that the above algorithm returns an integral solution of expected value at least $(3/4+c)\Opt$, for
some constant $c>0$, in which case we are done.

Otherwise, if the Non-Unique Prices Assumption is not true, we will remove all items in $N(\mu)$ from our fractional solution. Furthermore,
for the remaining items $j$, we will remove their assignment to players $i$ such that $p_{ij}\notin[(1-\mu)w_j,(1+\mu)w_j]$. These two
modifications imply that for our new fractional solution $x'$ it holds that 
\begin{align}
\sum\limits_{i,j} x'_{ij}p_{ij}\geq& (1-O(\mu,\epsilon_2))\sum\limits_{i,j} x_{ij}p_{ij}\nonumber\\
\geq& (1-O(\mu,\epsilon_1,\epsilon_2))\Opt \nonumber
\end{align}

Finally, in order to maintain that all players have full budgets, we scale down every player's budget appropriately; the LP value 
 of $x$ is unaffected by this modification.
 
 One thing we need to point out is that, even though in the following subsections we assume that every item has a unique price for all the players
 it can be assigned to, this is not exactly what we have enforced so far. Instead, we have enforced that the prices of an item for any two players it can be 
 assigned to may differ by a factor of at most $\frac{1+\mu}{1-\mu}$. The inaccuracy that is inserted to our analysis due to this discrepancy
 is analyzed in Section \ref{sec:wrap:non-unique}.
 
 This is a good place to remind the reader that, for our new fractional solution $x'$, it holds that all players have full budgets, and all items have 
 (almost) unique prices.

\subsection{Dealing with non-canonical solutions}\label{sec:wrap:canonical}
Next, we will ensure our fractional solution is canonical, by ensuring the following conditions for our updated fractional solution $x$: for
all $i\in\players$ and all
$j\in\bigitems_i$, $p_{ij}=B_i$
for all $i\in\players$, and for all $i\in\players$
$\sum\limits_{j\in\bigitems_i} x_{ij}p_{ij}=\sum\limits_{j\in\smallitems_i} x_{ij}p_{ij}=B_i/2$. 
In order to do so, let us introduce the concept of well structured solutions: a solution $x$ will be called 
$(\epsilon_3,\delta)$-well-structured, if at least an $1-\epsilon_3$ fraction of the LP value comes from players $i$ such that 
$\sum\limits_{j\in\bigitems_i} x_{ij}\in[(1-\delta)/2,(1+\delta)/2]$. Now, the following lemma is proved in  
\cite{DBLP:conf/ipco/KalaitzisMNPS14} (it appears as Lemma 5). Remember $\beta$ is the constant that defines the size of big items, and
that $x$ is a solution to the Assignment LP that is derived from a solution $y$ to the Configuration-LP (this fact is crucial in the proof
of the lemma): 

\begin{lemma}
\label{lem:non-wellstruct}
  Given a solution $x$ to the Assignment-LP which is not $(\epsilon_3, \delta)$-well-structured and
  $\beta \ge \delta / 4$, we can in polynomial time find a solution with expected value at least
  $\frac{3 + \epsilon_3 \delta^2 / 64}{4}\Val(x)$.
\end{lemma}

Therefore, whenever the conditions of Lemma \ref{lem:non-wellstruct} are satisfied, we can achieve an approximation guarantee strictly 
better than $3/4$ by choosing $\mu$, $\epsilon_1$ and $\epsilon_2$ to be much smaller than $\delta$ and $\epsilon_3$;in which case we would
be done. 

Otherwise, we know that $x$ is $(\epsilon_3, \delta)$-well-structured, and we do the following: 
first, we remove all players $i$ for which 
$\sum\limits_{j\in\bigitems_i} x_{ij}\notin[(1-\delta)/2,(1+\delta)/2]$ (which incurs a loss of a multiplicative factor of $(1-\epsilon_3)$
in the LP value). 

Next, in order to ensure that for all $i\in\players$ and all $j\in\bigitems_i$, $p_{ij}=B_i$, we will look at 
all $j\in \items$ and all $i:j\in\bigitems_i$, and set $p_{ij}=B_i$; this will result in a multiplicative loss of performance in any
subsequent rounding by a factor of at most $(1-\beta)$;
to see this, observe that this modification results in a loss in performance only if a big item is assigned to a player $i$. Then, 
such an item will have a fake value of $B_i$, but its original value will be at least $(1-\beta)B_i$; therefore, any guarantee we have on the 
performance of a rounding algorithm will have to be scaled by a factor of $1-\beta$.
% \sout{since in the worst case the rounding always picks a big item and then we lose a factor of
% $1-\beta$}
Notice that this modification might contradict the fact that we already forced an item $j$ to be assigned only to players whose
price for $j$ is close to $j$'s average price, but since we have control over the choice of $\beta$, we will see later on that we can deal
with this unpleasant side-effect.

In addition, in order to guarantee that $\sum\limits_{j\in\bigitems_i}x_{ij}=1/2$, we either decrease the assignment of every
big item uniformly, which would decrease the LP value by a multiplicative factor of $1-O(\delta)$, or insert fake big items of value $B_i$
into $i$; in this case, since all the rounding algorithms we use satisfy the marginals of the rounded solution for big items, we lose a
factor of $O(\delta)\Opt$ in the value of the final rounded solution.

Furthermore, in order to ensure that for all $i$ for all $i\in\players$, 
$\sum\limits_{j\in\smallitems_i}x_{ij}p_{ij}=B_i/2$, we will either lower the assignment of all small items assigned to $i$ by an
appropriate factor, which would imply that we lose an $O(\delta)$-fraction of our LP value, or insert one fake small item $j$ into $i$ with
$x_{ij}=1/2$ and $p_{ij}=2(B_i/2-\sum\limits_{j\in\smallitems_i}x_{ij}p_{ij})$ (which is at most $B_i/2$ if we choose $\delta$ to be small
enough; this is important because it allows us to invoke Lemma \ref{lem:ST2} in the following subsections), which implies that we have a loss in
performance upper
bounded by $O(\delta)\Opt$ in our final approximation guarantee To see this, observe that in the worst case this fake item is always picked, and $p_{ij}\in
O(\delta)$ as a result of the budgets being originally saturated and the definition of well-structured solutions.

In total, if $x'$ is the updated fractional solution after the above modifications, we know that 
\begin{align}
 \sum\limits_{i,j}x'_{ij}p_{ij}\geq& (1-O(\delta,\epsilon_3))\sum\limits_{i,j}x_{ij}p_{ij}\nonumber\\
 \geq&(1-O(\mu,\delta,\epsilon_1,\epsilon_2,\epsilon_3))\Opt\nonumber
\end{align}
while any rounding algorithm for $x'$ will have a loss in the expected returned value of at most $O(\delta)\Opt$ due to the insertion of
fake items.

To sum things up, at this point $x'$ satisfies the following properties: it is canonical, all players have full budgets, and all items have unique prices.

\subsection{Dealing with valuable small items}\label{sec:wrap:valuable-small}
Next, we will ensure our fractional solution satisfies the Cheap Small Items Restriction. In order to achieve this, let us assume first
that at least an $\epsilon_4$-fraction of the LP value of our updated solution $x$ comes from players for which the Valuable Small Items
Assumption is satisfied, i.e., from players $i$ such that 
 $$
 \sum\limits_{j:j\in\smallitems_i\wedge p_{ij}\geq B_i(1/2+\lambda)}x_{ij}\geq \epsilon_4
 $$
 for some small constants $\lambda,\epsilon_4>0$. Then, according to Lemma \ref{lem:ST2}, the ST rounding algorithm returns an integral
solution of expected value at least $(3/4+\Omega(\lambda,\epsilon_4))\sum\limits_{i,j} x_{ij}p_{ij}$; choosing $\mu$, $\delta$,
$\epsilon_1$, $\epsilon_2$ and $\epsilon_3$ to be much smaller than $\lambda$ and $\epsilon_4$, we get that the returned integral solution
has value at least
$(3/4+c)\Opt$, for some constant $c>0$. 

Otherwise, if at most an $\epsilon_4$-fraction of the LP value of our updated solution $x$ comes from players for which the Valuable Small
Items Assumption is satisfied, we remove all such players from our fractional solution, which incurs a loss of a multiplicative factor of
$1-\epsilon_4$ in our LP value. Furthermore, for any player $i$ and item $j\in\smallitems_i$ such that $p_{ij}>(1/2+\lambda)B_i$, we round
$p_{ij}$ down to $1/2$. In order to ensure our solution is still canonical, we insert a fake small item $j$ into $i$ with $x_{ij}=1/2$ and
appropriate $p_{ij}$ in order to ensure that $\sum\limits_{j\in\smallitems_i} x'_{ij}p_{ij}=B_i/2$. Then, any algorithm that rounds $x'$
will have a loss of a factor of $O(\epsilon_4,\lambda)\Opt$ in the returned integral solution, since in the worst case the fake small item
is always assigned, and its value is $O(\epsilon_4,\lambda)B_i$ due to the Valuable Small Items Assumption not being originally satisfied
for $i$. In total, we have
\begin{align}
 \sum\limits_{i,j}x'_{ij}p_{ij}\geq& (1-O(\epsilon_4))\sum\limits_{i,j}x_{ij}p_{ij}\nonumber\\
 \geq& (1-O(\mu,\delta,\lambda,\epsilon_1,\epsilon_2,\epsilon_3,\epsilon_4))\Opt\nonumber
\end{align}
while any algorithm that rounds $x'$ will have a loss of $O(\epsilon_4,\lambda)\Opt$ due to the insertion of fake items.

At this point, $x'$ has the following properties: it is canonical, all players have full budgets, all items have unique prices, and small items have
a relatively small price for any player they can be assigned to.

\subsection{Dealing with not fully assigned items}\label{sec:wrap:fully-assigned}
Now, let us assume that a significant fraction of the total LP value comes from real items $j$ such that
$x_j<9/10$, i.e., that
\begin{assumption:not-full-items}
 $$\sum\limits_{i\in\players}\sum\limits_{j\in
\items:x_j<9/10}x_{ij}p_j\geq \epsilon_5\sum\limits_{i\in\players}\sum\limits_{j\in
\items}x_{ij}p_j$$
for some small constant $\epsilon_5>0$.
\end{assumption:not-full-items}

Then, by scaling up the assignment of every such item to every player, and using the facts that $x$ was originally canonical and Lemma
\ref{lem:ST3}, we get that choosing $\mu$, $\delta$, $\lambda$, $\epsilon_1$, $\epsilon_2$, $\epsilon_3$ and $\epsilon_4$ to be much
smaller than $\epsilon_5$, the ST rounding algorithm returns a solution of expected value at least $(3/4+c)\Opt$.

On the other hand, if the Not Fully Assigned Items Assumption is not satisfied, we can remove all items $j$ for which $x_j<9/10$ from our
fractional solution, and fill up the budget of every player with an appropriate amount of fake big and small items such that to ensure that
the produced solution $x'$ remains canonical. Then, due to the insertion of fake items, we get that any algorithm that rounds $x'$ will
lose a factor of $O(\epsilon_5)\Opt$ in the final integral solution value. Therefore, we have that 
\begin{align}
 \sum\limits_{i,j}x'_{ij}p_{ij}\geq& (1-O(\epsilon_5))\sum\limits_{i,j}x_{ij}p_{ij}\nonumber\\
 \geq&(1-O(\mu,\delta,\lambda,\epsilon_1,\epsilon_2,\epsilon_3,\epsilon_4,\epsilon_5))\Opt\nonumber
\end{align}
and that any algorithm that rounds $x'$ will have a loss of $O(\epsilon_5)\Opt$ in the final returned value due to the insertion of fake
items.

Finally, observe that here we ensured that all real items are almost fully assigned, while in Section \ref{sec:nup} we made that assumption for
all items. The only place where this discrepancy makes a difference is Lemmas \ref{lem:nup:bi} and \ref{lem:nup:Na}.
However, since for any player the total
fractional assignment of fake items is $O(\mu,\delta,\lambda,\epsilon_1,\epsilon_2,\epsilon_3,\epsilon_4,\epsilon_5)$, picking these
constants to be small enough, these lemmas still hold.

From now on, $x'$ satisfies the following: it is canonical, all players have full budgets, all items have unique prices, small items have
a relatively small price for any player they can be assigned to, and all items are almost fully assigned.

\subsection{Dealing with fake items and non-unique prices}\label{sec:wrap:non-unique}
At this point, we should pause for a moment and make the following observation:  fake items might be scaled up during the execution of
the Big-Small Items Algorithm. This means that we should restrict the
Big-Small Items Algorithm to only operate on real items. The only places where our analysis
will be imprecise because of this discrepancy, is in Lemmas \ref{lem:nup:bi} and \ref{lem:nup:Na}. However, since for any player the total
fractional assignment of fake big items is $O(\mu,\delta,\lambda,\epsilon_1,\epsilon_2,\epsilon_3,\epsilon_4,\epsilon_5)$, picking these
constants to be small enough, these lemmas still hold.
 
 Finally, during the analysis of the same algorithm, we assumed all items have the same price, while in reality two prices of an item might
vary by a factor of $O(\mu,\beta)$ (the $\mu$ factor comes from dealing with the Non-Unique Prices Assumption, and the $\beta$ factor from
ensuring that $x$ is canonical); since the Big-Small Items Algorithm shifts the assignment of items, we might lose up to an $O(\mu,\beta)$
factor
of the LP value of every item, which results in the loss of $O(\mu,\beta)\Opt$ in our final rounding performance. Then, picking $\mu$ to be
small
enough, Lemma \ref{lem:nup:Na} will still hold.

\subsection{Proof of Theorem \ref{thm:main}}\label{sec:wrap:main}
As is shown in Section \ref{sec:nup}, when the Big-Small Items Assumption is true, i.e., when at least an $\epsilon_6$-fraction of the LP
value comes from items in $M(\nu)$, then there exists a rounding algorithm which achieves an approximation guarantee strictly better than
$3/4$. 
Therefore, we assume the assumption is false. Then, we will remove all items in $M(\nu)$ from our fractional solution $x$, and fill the budget of all players with
fake
big or small items, according to whether the total assignment of big or small items to some player should be increased; this step will
result in the loss of a factor of $\epsilon_6\Opt$ in the performance of any rounding algorithm. Similarly, for any item $j$ which is not in $M(\nu)$, we restrict it to be assigned only to players that value it as big, if it assigned mostly as big, or to players that value it as small, and fill up the budgets of all players accordingly. This step results in the loss of a factor of $\nu\Opt$ in the performance of any rounding algorithm we will see from now on.

To sum things up, $x'$ finally satisfies the following properties:
\begin{itemize}
 \item It is canonical.
 \item All players have full budgets.
 \item All items have unique prices.
 \item Small items have relatively small prices for any player they can be assigned to.
 \item All items are almost fully assigned.
 \item All items are either assigned exclusively as big or exclusively as small.
\end{itemize}

Now, the work of Kalaitzis et al. \cite{DBLP:conf/ipco/KalaitzisMNPS14} gives the following result (the original result is more general, and requires less
restrictions, but this version is tailored to our purposes):
\begin{theorem}\label{thm:ipco}
 Let $y$ be a fractional solution to the Configuration-LP, and let $x$ be a fractional solution to the Assignment-LP, derived  from $y$,
which satisfies the following properties:
 \begin{itemize}
 
  \item[(a)] for all $i\in\players$ and all $j\in\bigitems_i$, $p_{ij}=B_i$.
  \item[(b)] for all $i\in\players$, $\sum\limits_{j\in\bigitems_i} x_{ij}p_{ij}=\sum\limits_{j\in\smallitems_i} x_{ij}p_{ij}=B_i/2$.
  \item[(c)] for all $j\in\items$, $x_{ij}>0$ implies $p_{ij}=p_j$, for some $p_j\geq 0$; of course, different items are allowed to have
different $p_j$.
  \item[(d)] for all $j\in\items$, $\sum\limits_{i:j\in\bigitems_i} x_{ij}=0$ or $\sum\limits_{i:j\in\smallitems_i} x_{ij}=0$.
 \end{itemize}
Then, there exists a randomized rounding algorithm, which rounds $x$ to an integral solution, such that the expected value of the returned
integral solution is at least $(3/4+c)\sum\limits_{i,j}x_{ij}p_{ij}$, for some constant $c>0$, and
\begin{itemize}
 \item the probability of big item $j$ being assigned to $i$ is equal to $x_{ij}$.
 \item the probability of small item $j$ being assigned to $i$ is at most $2x_{ij}$.
\end{itemize}
\end{theorem}

Finally, our LP
solution $x'$ obeys the conditions of Theorem \ref{thm:ipco}, which already
provides a larger than $3/4$-approximation guarantee. 
Since the marginals of big items are satisfied, and since the marginals of small items are satisfied up to a
constant factor, the total value of fake items in the final integral solution is 
$O(\mu,\nu,\delta,\lambda,\epsilon_1,\epsilon_2,\epsilon_3,\epsilon_4,\epsilon_5,\epsilon_6)$. Therefore, picking 
$\mu$, $\nu$, $\delta$, $\lambda$, $\epsilon_1$, $\epsilon_2$, $\epsilon_3$, $\epsilon_4$, $\epsilon_5$ and $\epsilon_6$ to be much
smaller than $c$,  Theorem \ref{thm:main} follows.
\balance
\bibliography{mba}

\end{document}